\documentclass[useAMS,usenatbib,referee]{mn2e}
\newcommand{\case}{$\mathrm{case}_{45}$}
 
\usepackage{epsfig}
\usepackage{graphicx}
\title[A two-dimensional
electrodynamical study on the gamma-ray pulsar]{A two-dimensional
electrodynamical outer gap model for gamma-ray pulsars:Gamma-ray spectrum}
\author[J.~Takata, S.~Shibata, K.~Hirotani \& H.-K.~Chang]
  {J.~Takata,$^1$\thanks{takata@tiara.sinica.edu.tw}
  S.~Shibata,$^2$ K.~Hirotani,$^1$ H.-K.~Chang$^3$ \\
  $^1$ASIAA/National Tsing Hua University - TIARA, Hsinchu, Taiwan\\
  $^2$Department of Physics, Yamagata University, Yamagata 990-8560,
Japan \\
$^3$Department of Physics and Institute of Astronomy,
 National Tsing Hua University, Hsinchu 30013, Taiwan}

\date{Released 2005 Xxxxx XX}

\def\LaTeX{L\kern-.36em\raise.3ex\hbox{a}\kern-.15em
    T\kern-.1667em\lower.7ex\hbox{E}\kern-.125emX}

\begin{document}

\label{firstpage}

\maketitle

\begin{abstract}
A two-dimensional electrodynamical model is used to study 
particle acceleration in the  outer magnetosphere of a pulsar.
 The charge depletion from the
Goldreich-Julian charge density causes a large electric field  along
 the  magnetic field lines.  The charge particles are accelerated by the
 electric field and emit $\gamma$-rays via the  curvature
process. Some of the emitted $\gamma$-rays may collide with 
$X$-ray photons to make new pairs, which are accelerated again on the
different field lines in the gap and proceed similar processes. 
We simulate the pair creation 
cascade in the meridional plane using the pair creation mean-free
path, in which the $X$-ray photon number density is proportional to inverse
square of radial distance.   
With the space charge density determined by  the pair creation simulation, 
we solve the electric  structure of the outer gap in the meridional
plane  and  calculate the curvature spectrum. 

We investigate in detail relation
between the spectrum and total current, which is carried by the
particles produced in the gap and/or injected at the boundaries of the gap. 
We demonstrate that  the hardness of the spectrum  is strongly
controlled by the current carriers. Especially, the spectrum sharply
softens if we assume a larger particle injection 
at the outer boundary of the outer gap.  This is because  
the mean-free path of the pair creation of the inwardly propagating 
$\gamma$-ray photons is much shorter than the light radius so that the many
pairs are produced in the gap to quench the outer gap.

Because the two-dimensional model 
can link  both  gap width  along the magnetic field line and 
trans-field thickness with the spectral cut-off
energy and flux, we can diagnose both the current through 
 the gap and  inclination angle between the
rotational and magnetic axes. We apply the theory to the Vela
pulsar. By comparing the results with the $EGRET$ data, 
we rule out  any cases that have  a large particle injection at the outer
boundary. We also  suggest the inclination angle of
$\alpha_{inc}\geq$65$^{\circ}$.
 The present model  predicts the outer gap starting from near the
conventional null charge surface for the Vela pulsar.

\end{abstract}
\section{Introduction}
Rapidly rotating young pulsars such as the Crab and  Vela pulsars emit
high-energy $\gamma$-rays and are  among  the  brightest sources 
in the $\gamma$-ray sky. In the pulsar magnetosphere, electrons and/or
positrons are accelerated above 10~TeV. Because the pulsar is an electric 
dynamo with an available potential drop,
 $\Phi_{a}\sim 1.3\times 10^{13}P^{-3/2}\dot{P}^{1/2}_{-15}$ Volt,
where $P$ is the rotational period and $\dot{P}_{-15}$ is the 
time derivative of the rotational period in units of $10^{-15}$ s/s, 
 theoretically, it is suggested that the pulsars utilize some fraction of 
$\Phi_a$ to accelerate the particles.
 \textit{EGRET} instrument observed  $\gamma$-ray 
emissions from 7 pulsars (Thompson et al. 1999). 
The observed modulation of light curves and spectral cut-off
energy  test the theoretical models; e.g.,  
the polar cap accelerator model (Sturrock 1971; Ruderman \& Sutherland
1975; Scharlemann, Arons \& Fawley 1978; Daugherty \& Harding 1996),  
the outer gap accelerator model (Cheng, Ho \& Ruderman 1986a,b,
hereafter CHR86; Romani 1996; Zhang \& Cheng 1997; Hirotani \& Shibata
1999a,b), and 
the inner annual gap accelerator model (Qiao et al. 2004).
 All models predict the particle  acceleration in a charge
depletion region called the gap, in which the charge density differs from the
Goldreich-Julian charge density, with an  electric field parallel to the
local magnetic field lines. 

The outer gap model has been successful in explaining the  observed
main features of the $\gamma$-ray light curves, such as the two peaks in a
period and the presence  of the  emissions between the peaks. 
 CHR86 first investigated electric structure of the outer gap. 
 By solving the Poisson 
equation for the  vacuum gap,  CHR86 showed that the inner boundary of the
vacuum gap  locates close to  the null charge surface 
of the Goldreich-Julian charge density, and concluded that
 the outer gap is lying above the last open lines  and  is
 extending between the null charge surface  
and the light cylinder, the axial distance of which is
  $\varpi_{lc}=c/\Omega$, where $\Omega$ is the 
rotational angular frequency and $c$ the speed of light. 
The last open lines are defined by the magnetic field lines 
tangent to the light cylinder. With CHR86's 
vacuum gap geometry, the two peaks in a period have been  interpreted
as an effect of aberration and time delay of the emitted photons
(Romani \& Yadigaroglu 1995, hereafter RY95; Cheng, Ruderman \& Zhang
2000). RY95 also reproduced phase separation of the pulses in different
 photon energies from radio to $\gamma$. 

Hirotani \& Shibata (1999a,b, hereafter HS99) have  solved 
electrodynamics in the non-vacuum outer gap  (see also
 Hirotani \& Shibata 2001a,b).  
Although HS99 worked in only one dimension along the last open  line, they  
 first  connected consistently  the field-aligned electric field 
in the outer gap with the curvature radiation and pair creation
processes. The model  reproduced the  $EGRET$ phase-averaged spectra 
of the Vela like pulsars.

With these successful works, the outer gap has been considered 
as a promising acceleration model accounting for the high energy
emissions of the pulsars. However, there are some problems in
the previous works. For example,  CHR86 discussed the electrodynamical
structure of  the vacuum gap, although they assumed an exponential grow of
the number density  in the trans-field direction. Such a large number
of particles
must partially screen the field-aligned electric field on the magnetic
field lines, on which they flow. Another problem is the
one-dimensionality in the HS99 arguments. 
The one-dimensional model  neglected  important trans-field effects 
caused by the curvature of the field
lines. The $\gamma$-rays are emitted 
tangent to the local magnetic field lines, 
converts into pairs on different field lines, 
and as a result, makes a trans-field distribution of the current. It is 
obvious that the electrodynamics such as  the strength of the
 field-aligned  electric field  is affected by the trans-field
 structures. Furthermore, the one-dimensional electrodynamical model has
used the trans-field thickness as a parameter to be determined by 
the observed flux (Takata, Shibata \& Hirotani 2004a, hereafter Paper1).

Recently, the emission region extending to both the stellar surface
and the light cylinder has been proposed to explain the presence of  
the observed outer-wing emissions and of the off-pulse emissions of the Crab pulsar 
(Dyks \& Rudak 2003; Dyks, Harding \& Rudak 2004).  Muslimov \&
Harding (2004) proposed the  \textit{slot gap} accelerator model (Arons 1983), 
which bases on the polar cap model, as an  
origin of the emission region by  Dyks \& Rudak (2003). 
On the other hand, it has been  suggested that the outer gap 
may have an elongated emission regions (Takata, Shibata \& Hirotani
2004b, hereafter Paper2). Thus, the
origin of the $\gamma$-ray emissions in the pulsar magnetosphere has
not been conclusive up to now. 

In Paper2, we have  solved  the
electrodynamics in the outer gap with a  two-dimensional 
model in the meridional plane. 
We extended HS99's one-dimensional model into two-dimensional one
with the trans-field structure. This two-dimensional electrodynamical
model links the one-dimensional
electrodynamical outer gap model by HS99 and the three-dimensional geometrical
outer gap model by RY95.

 In Paper2, we showed  that 
 the cusp of the inner boundary is located at the position where 
the local charge density caused by the current is equal to the local
Goldreich-Julian value.
This implies that the inner boundary of the outer gap shifts toward
stellar surface from the conventional null charge surface as the
current increases. 
In Paper2, the elongated gap  was predicted  by  the 
cases of the particle injection at the outer boundary of the
gap. Thus, the outer gap indicates the acceleration region 
predicted by Dyks \& Rudak (2003).
 However, because the spectrum was not investigated in Paper2, 
it was not certain that what kind of gap structure  reproduces 
the observed $\gamma$-ray spectrum.

The purpose of this paper is to calculate the $\gamma$-ray spectrum
 using the two-dimensional electrodynamical model proposed 
in Paper2 and to examine  dependence of the spectrum on the current
through the gap  and the inclination angle between the rotational and
magnetic axes. We can estimate flux as seen
from the Earth more correctly than the one-dimensional model, 
because we solve the gap structure in 
 both longitudinal and trans-field directions. 
We discuss the particle acceleration in the pulsar magnetosphere, 
and constrain the model parameters  in terms of the 
spectral cut-off energy and the flux. To compare with the observations, 
we relax  the previous restrictions of the model such as 
\begin{itemize}
\item the inclination  angle is equal to zero (i.e. aligned-rotator), 
\item  $X$-ray field of the pair creation is homogeneous in the
magnetosphere, 
\item  all $\gamma$-rays produced by  the curvature process have
critical energy.
\end{itemize}   
In this paper, we calculate the pair creation process with 
inhomogeneous and anisotropic  $X$-ray
field and the curvature spectrum of the individual particles
(\S\S\ref{casmodel}). 

We demonstrate that the calculated spectrum softens if one assumes a larger 
current through the gap. Especially, the calculated spectrum 
sharply softens with the increasing the number of  particles injected 
at the outer boundary of the outer gap. On the other hand, 
 we also demonstrate that an increase in the number of particles injected 
 at the inner boundary or of the inclination angle  moderately 
changes  the calculated spectral hardness. Using this dependence of
the spectral hardness, therefore, we can investigate the current  
in the outer gap of the observed $\gamma$-ray pulsars 
by comparing   the model spectrum with the observed one. 
We apply the theory to the Vela pulsar, 
and show the gap geometry, the current through the gap and the inclination 
angle ($\equiv\alpha_{inc}$) in the Vela magnetosphere.  We show that 
 the  $EGRET$ spectrum predicts a large inclination angle
$\alpha_{inc}\geq65^{\circ}$ and only a few particle 
injections into the outer gap at the outer boundary.  

In \S\ref{beq}, we present the  basic equations describing 
the stationary outer gap
structure. In \S\S\ref{casmodel}, we develop the  previous pair creation
cascade model in Paper2  using the Monte Carlo method. 
In \S\ref{results}, we compare the
results with the observation of the Vela pulsar. In this paper, 
the polarity is $\mathbf{\Omega}\cdot\mathbf{m}\geq0$, where
$\mathbf{m}$ 
is the magnetic moment of the star. Since we ignore the effects of ions, the
results do not depend on the polarity. 

\section{Model \& Basic equations}
\label{beq}
For solving  a stationary structure of the outer gap, we treat the ideal
conditions:
\begin{enumerate}  
\renewcommand{\theenumi}{$Con$\arabic{enumi}}  
\item the stationary condition ($\partial_t+\Omega\partial_{\phi}=0$,
where $t$ and $\phi$ are the time and azimuth, respectively) is
satisfied, that is, there is no time variation of any quantities as
seen in a co-rotating system,
\item the star crust is a rigid rotating perfect conductor,
\item the plasmas fill the pulsar magnetosphere, and the condition 
$\mathbf{E}\cdot\mathbf{B}=0$ holds all the way along the field lines
connecting the star and the outer gap accelerator.
\end{enumerate}
Strictly speaking, the condition $Con3$ cannot arise in the first
place. We assume that a mechanism of the electric
field screening  works in the region between the stellar surface and 
the inner boundary of the outer gap. In \S\S\ref{inpart}, 
we discuss the dynamics outside the gap.

By neglecting the magnetic component by  rotational effects 
and by the  current, we assume the static dipole magnetic field. 
This static dipole structure will be a good approximation as long as  
the electrodynamics in the outer gap is discussed. 
In fact, the sweepback effect is important in the
geometrical studies to produce the double peaks 
in the light curves (Cheng, Ruderman \& Zhang 2000).
Electrodynamical speaking,  the magnetic field affects 
on the electrodynamics thorough the Goldreich-Julian charge density, 
that is, the magnetic component projected to the rotational axis.
 As compared with the static dipole field, the null charge surface of
the Goldreich-Julian charge density on the last-open lines, and therefore 
 the inner boundary of the outer gap for the rotating dipole field  
is located  closer to the stellar surface about 10\% of the light radius.  
With this small difference of the positions of the  inner boundaries
between two magnetic fields, the qualitative arguments on the
electrodynamics in the outer gap will  not be changed 
very much  as long as  the outer boundary of the gap is not located
near the light cylinder. In the later sections,  we will 
predict that the out boundary that is favored for the
Vela spectrum  does not locate near the light cylinder (Fig.\ref{allgap}), 
 while the outer boundary on the light cylinder has been assumed 
in the previous light curve models (Romani 1996; Cheng, Ruderman \&
Zhang 2000). It is also worth noting  that even if the outer boundary 
is located inside of the light cylinder, the two peaks in the light curves are 
expected because the outward moving particles accelerated in the gap
 continue to emit the high energy photons outside the gap  
via the curvature process (\S\S\ref{compari}). 

\subsection{Poisson equation}
\label{basicpoi}
With the condition $Con1$, the electric field can be written as 
(Shibata 1995, Mestel 1999)
\begin{equation} 
\mathbf{E}(\mathbf{r},t)=-(\Omega\mathbf{e}_{z}\times\mathbf{r})
\times\mathbf{B}/c-\nabla\Phi_{nco}(\mathbf{r},t),
\label{geelec} 
\end{equation} 
where $\Phi_{nco}$ is the non-corotational potential and $\mathbf{e}_{z}$
is the unit vector along the rotation axis. 
The $E_{||}$-acceleration is indicated by the non-corotational part in
equation (\ref{geelec}),
$E_{||}=-\mathbf{B}\cdot\nabla\Phi_{nco}/B\neq0$.  
Substituting equation (\ref{geelec}) in $\nabla\cdot\mathbf{E}=4\pi
\rho$ gives  the Poisson equation, 
 \begin{equation}  
\triangle\Phi_{nco}(\mathbf{r})=-4\pi[\rho(\mathbf{r}) 
-\rho_{GJ}(\mathbf{r})],  
\label{poisson}  
\end{equation}  
where $\rho(\mathbf{r})$ is the space charge density, and
$\rho_{GJ}(\mathbf{r})$ is the Goldreich-Julian charge density, 
\begin{equation}  
\rho_{GJ}=-\left(\frac{\Omega}{2\pi c}\right)\mathbf{e}_z\cdot  
\left[\mathbf{B}-\frac{1}{2}\mathbf{r}\times(\nabla\times\mathbf{B})\right].
\label{rhogj}
\end{equation}
To simplify the geometry, we assume that the gap dimension in 
the azimuthal direction is much larger than that in the meridional
plane. Neglecting  variation in the azimuthal direction, 
we rewrite  equation (\ref{poisson}) as 
\begin{equation} 
\triangle_{r,\theta}\Phi_{nco}(\mathbf{r})
=-4\pi[\rho(\mathbf{r})-\rho_{GJ}(\mathbf{r})],  
\label{basic1} 
\end{equation}
 where $\triangle_{r,\theta}$ represents ($r,\theta$)-parts of 
 the Laplacian. 

At the moment, let's rewrite down the Poisson equation (\ref{basic1})
as 
\begin{equation}
\frac{\partial^2\Phi_{nco}}{\partial
s_{||}^2}+\frac{\partial\Phi_{nco}}{\partial s_{\perp}^2}
\sim-4\pi(\rho-\rho_{GJ}),
\label{simpoi}
\end{equation}
 where $s_{||}$ and $s_{\perp}$ are the distance along the magnetic
field  and trans-field directions, respectively. Near the boundaries of 
the gap or  for a thick outer gap, in which typical gap width along
the filed line is much larger than the typical trans-field 
thickness, the left hand side of equation (\ref{simpoi}) is dominated by 
 the first term. The accelerating
electric field is determined by   
\begin{equation}
E_{||}=-\frac{\partial \Phi_{nco}}{\partial s_{||}}\sim 
\int4\pi(\rho-\rho_{GJ})ds_{||}.
\end{equation} 
 Apart from the boundaries for the thin outer gap, on the other hand, 
the second term  dominates in the left hand side of  equation 
(\ref{simpoi}). Hence, the accelerating
electric field behaves such as (CHR86)
\begin{equation}
E_{||}\sim2\pi s_{\perp}(s_{\perp}-D_{\perp})
\frac{\partial}{\partial s_{||}}(\rho-\rho_{GJ}),
\end{equation}
where we ignore the trans-field dependence of the charge density, 
$D_{\perp}$ is the typical gap trans-field thickness, and
$s_{\perp}=0$ and $s_{\perp}=D_{\perp}$  represent the lower and
upper boundaries of the gap, respectively. We find that the magnitude of the 
 gradient of the  effective charge
density ($=\rho-\rho_{GJ}$) along the field lines determines the
strength of the field-aligned electric field for the thin gap geometry,
although the magnitude of the effective charge density causes the
field-aligned electric field for the thick gap geometry. In a real pulsar
magnetosphere, the thin  gap or the thick gap 
will be controlled by the pair-creation mean free path. 

To solve the Poisson equation (\ref{basic1}) numerically, 
we introduce an orthogonal curvilinear coordinate system
($\chi,\zeta$) following the dipole field on the meridional plane as 
\begin{equation}  
\chi\equiv\frac{r\sin^{-2}(\theta-\alpha_{inc})}{\varpi_{lc}} 
=\textrm{constant along a dipole   
field line},  
\end{equation}  
and     
\begin{eqnarray}  
\zeta\equiv\frac{r\cos^{-1/2}(\theta-\alpha_{inc})}{\varpi_{lc}}&= 
&\textrm{constant along a curved line}  
\\ \nonumber  
&& \textrm{perpendicular to the field lines},  
\end{eqnarray}
where $r$ and $\theta$ is the radial distance and the colatitude angle
with respect to the rotational axis, respectively.  The Laplacian
becomes 
\begin{equation} 
\triangle_{r,\theta}=\frac{3\cos^2\theta'+1}{\sin^6\theta'}\frac{\partial^2}
{\partial \chi^2}+\frac{4}{\chi\sin^6\theta'}\frac{\partial}{\partial
\chi}  
+\frac{3\cos^2\theta'+1}{4\cos^3\theta'}\frac{\partial^2}{\partial
\zeta^2}  
+\frac{3(3\cos^2\theta'+1)}{4\zeta\cos^3\theta'}\frac{\partial}{\partial
\zeta}, 
\end{equation} 
where $\theta'\equiv \theta-\alpha_{inc}$.  Non-corotational electric
field ($\equiv\mathbf{E}_{nco}$) is given by 
\begin{equation}  
E_{||}=-\frac{\partial\Phi_{nco}}{\partial s_{||}}=  
-\frac{\sqrt{3\cos^2\theta'+1}}{2\varpi_{lc}\cos^{3/2}\theta'}  
\frac{\partial\Phi_{nco}}{\partial \zeta},  
\end{equation}  
\begin{equation}  
E_{\perp}=-\frac{\partial\Phi_{nco}}{\partial s_{\perp}}=  
-\frac{\sqrt{3\cos^2\theta'+1}}{\varpi_{lc}\sin^3\theta'}  
\frac{\partial\Phi_{nco}}{\partial \chi},  
\end{equation}  
where $ds_{||}=(2\cos^{3/2}\theta'/\sqrt{3\cos^2\theta'+1})d\zeta$ and
$ds_{\perp}=(\sin^3\theta'/\sqrt{3\cos^2\theta'+1})d\chi$ are the line
elements along the field line and the perpendicular 
curved line, respectively.

 \subsection{Particle continuity equations and pair creation process}
\label{contsec}
For the young pulsar such as the Vela pulsar, the non-corotational drift 
motion of the charged particles due to the non-corotational 
electric field $\mathbf{E}_{nco}$ is negligibly small in comparison 
with  the corotational motion
(Paper2). Also ignoring the gyro-motion, we denote 
the velocity $\mathbf{v}$ as   
\begin{equation}  
\mathbf{v}=v_{||}\mathbf{e}_{||}+\varpi\Omega\mathbf{e}_{\phi},  
\end{equation}  
where $v_{||}$ represents the longitudinal velocity.
 Because the accelerated particles move at the speed of light, the 
longitudinal velocity in the meridional plane is  to be 
$v_{||}=\sqrt{c^2-(\varpi\Omega)^2}$, with which 
the particles migrate along the line of $\chi=\mathrm{constant}$. 

With $Con1$, the continuity equation yields 

\begin{equation}  
 \mathbf{B}\cdot\nabla\left(\frac{v_{||}N_{\pm}(\mathbf{r})}{B}\right)=\pm
S(\mathbf{r}),  
\label{basic2}  
\end{equation}  
where $S(\mathbf{r})$ is the source term,  and  $N_+$ and $N_-$  denote 
the number density of outwardly and inwardly moving particles
 (i.e. the positrons and the electrons), respectively.

To  calculate the  source term $S(\mathbf{r})$, 
we adopt the pair creation process between background 
 $X$-ray photons and the  $\gamma$-ray photons in the gap. 
The mean free path of a
 $\gamma$-ray photon with energy $E_{\gamma}$ is written  as 
\begin{equation} 
l_p(\mathbf{r},E_{\gamma})=\frac{c}{\eta_p(\mathbf{r},E_{\gamma})},
\label{meanfree} 
\end{equation} 
with 
\[  
\eta_p(\mathbf{r},E_{\gamma}) 
=(1-\cos\theta_{X\gamma})c\int_{E_{th}}^{\infty}dE_{X}\frac{dN_X}  
{dE_X}(\mathbf{r},E_{X})\sigma_p(E_{\gamma},E_{X}),  
\] 
where $dE_{X}\cdot dN_X/dE_{X}$ is the $X$-ray number  
density between energies $E_{X}$ and  
$E_X+dE_X$, $\theta_{X\gamma}$ is the collision angle  
between an $X$-ray photon and a $\gamma$-ray photon,  
$E_{th}=2(m_ec^2)^2/(1-\cos\theta_{X\gamma})E_{\gamma}$ is the
threshold $X$-ray energy for the  pair creation,  
and  $\sigma_p$ is the pair creation cross-section, which is given by
\begin{equation}  
\sigma_{p}(E_{\gamma},E_{X})=\frac{3}{16}  
\sigma_{T}(1-v^2)\left[(3-v^4)\ln\frac{1+v}{1-v}-2v(2-v^2)\right],  
\label{cross}  
\end{equation}  
where   
\[  
v(E_{\gamma},E_{X})=\sqrt{1-\frac{2}{1-\cos\theta_{X\gamma}}\frac{(m_ec^2)^2}
{E_{\gamma}E_{X}}},  
\]  
and  $\sigma_{T}$ is the Thomson cross section.  In the latter
sections, we apply the theory to the Vela pulsar, for which 
the observed $X$-ray spectrum is dominated by thermal emission. 
This component has been interpreted as the emissions from
the stellar surface. At the radial distance $r$ from the centre of the star, 
 the thermal photon number density between energy
$E_{X}$ and $E_{X}+dE_{X}$ is given by 
\begin{equation}
\frac{dN_X}{dE_X}=\frac{1}{4\pi}\left(\frac{2\pi}{ch}\right)^3
\left(\frac{R_{eff}}{r}\right)^2
\frac{E_X^2}{\exp(E_X/kT_s)-1},
\label{soft}
\end{equation}
where $R_{eff}$ is the effective radius of the emitting region, and
$kT_s$ refers to the surface temperature. For  the values of $R_{eff}$ and
$T_s$, the  observed ones are used. 

With the soft photons from the stellar surface, 
the collision angle $\theta_{X\gamma}$  of  the $\gamma$-ray photon 
after traveling the distance $s$ from the emission point $(r_0,\theta_0)$
is obtained from 
\begin{equation}
\cos\theta_{X\gamma}(\mathbf{r})=\frac{s+r\cos\theta_{em}}{r},
\label{coliang}
\end{equation}
where $\theta_{em}$ is the angle between the emission direction and the radial
direction at the emission point, which are 
$\cos\theta_{em}^{+}=\sqrt{1-(\varpi_0\Omega/c)^2}
B_r(r_0,\theta_0)/B(r_0,\theta_0)$ for the outwardly propagating
$\gamma$-rays,  and $\cos\theta_{em}^{-}=-\sqrt{1-(\varpi_0\Omega/c)^2}
B_r(r_0,\theta_0)/B(r_0,\theta_0)$ for the inwardly propagating
$\gamma$-rays.  For $\alpha_{inc}=45$$^{\circ}$, for example,
the $\gamma$-rays are emitted outward (or inward) from the
conventional null charge point on the last open line with
$\theta^+_{em}\sim 22.8$$^{\circ}$ (or $\theta^-_{em}\sim
157.2$$^{\circ}$).

\subsection{Curvature radiation process}
We calculate  the curvature radiation process of the
accelerated particles. The power per unit energy emitted by the individual 
electrons (or positrons) is written as 
\begin{equation}  
P_c(R_c,\Gamma,E_{\gamma})=  
\frac{\sqrt{3}e^2\Gamma}{hR_c} 
F(x),  
\label{curate} 
\end{equation}   
where $x\equiv E_{\gamma}/E_c$,  
\begin{equation}  
E_c=\frac{3}{4\pi}\frac{hc\Gamma^3}{R_c},  
\label{critic} 
\end{equation}  
and  
\begin{equation}  
F(x)=x\int_x^{\infty}K_{5/3}(t)dt,  
\end{equation}  
where $R_{c}$ is the curvature radius of the magnetic field line,
$\Gamma$  is the Lorentz factor of the particles,
 $K_{5/3}$ is the modified Bessel function of the  order 5/3, 
$h$ is the Planck constant, and $E_{c}$   
gives the characteristic curvature photon energy. The Lorentz 
factor in the gap is given by assuming that the particle's 
motion immediately saturates in the    
balance between the electric and the radiation reaction forces,   
\begin{equation}   
\Gamma_{sat}(R_c,E_{||})=\left(\frac{3R_c^2}{2e}E_{||}+1\right)^{1/4}.
\label{gamma}   
\end{equation}      

As discussed in Paper1, the curvature emissions of the  
outwardly moving particles from 
outside the gap also contribute to the total emissions if 
the outer boundary of the gap is not  located  close to the light cylinder.  
Outside the gap, the particles loose their energy via the 
curvature emissions so that 
\begin{equation}  
m_ec^2\frac{d\Gamma(s_{||})}{ds_{||}}=-\frac{2}{3}\frac{e^2\Gamma^4(s_{||})} 
{R_{c}^2(s_{||})}.  
\label{gameq}  
\end{equation}  
We assume that the particles migrating on 
 the magnetic field line labeled by $\chi$ escape  from the gap 
with a Lorentz factor given  by the value at $W_{||}(\chi)/c=t_d$, 
which describes the condition that the gap crossing time
$W_{||}(\chi)/c$ of the particles is equal to the radiation damping time 
($\equiv t_d$) estimated by  
\begin{equation}  
t_{d}\sim 4\cdot10^{-3}\left(\frac{\Omega}  
{100\mathrm{rads^{-1}}}\right)^{-1}  
\left(\frac{\Gamma}{10^7}\right)^{-3}  
\left(\frac{R_c}{0.5\varpi_{lc}}\right)^2~~s.   
\end{equation}  
In the present paper, we take into account  the curvature emissions up
to $\varpi=0.9\varpi_{lc}$.

The saturation (\ref{gamma}) in the gap simplifies the problem significantly.
The saturation motion of the particles will be achieved  if the 
the typical acceleration length scale $l_{ac}\sim
10^7(\Gamma/10^{7})(E_{||}/3\cdot10^7\mathrm{V\cdot m^{-1}})^{-1}$~cm 
is shorter  
than the gap width $W_{||}(\chi)$ (see
Paper1). For some part of the gap, where the electric field is week, 
 the saturation approximation will break down (Hirotani, Harding \&
Shibata 2003). In the two-dimensional model, specifically, 
around the lower and the upper boundaries,  the condition  $W_{||}>l_{ac}$ 
is not satisfied effectively for all  particles. 
Nevertheless, we adopt the Lorentz factor $\Gamma_{sat}$  
to all particles in the outer gap for simplicity,  because 
the saturation  breaks down within a few percent of the gap
thickness measured from  the upper and the lower boundaries and because 
the contributions of the emissions from such region 
 are less important for the observed spectrum above 100MeV. 

The high-energy particles also emit high-energy photons via the 
 synchrotron process and the inverse-Compton process. In
\S\S\ref{emipro},  we show that the both emission processes 
will be less important for the gap electrodynamics in comparison to
the curvature process.

\subsection{Pair creation position}
\label{casmodel}

\begin{figure}   
\begin{center}   
\includegraphics[width=8cm, height=7cm]{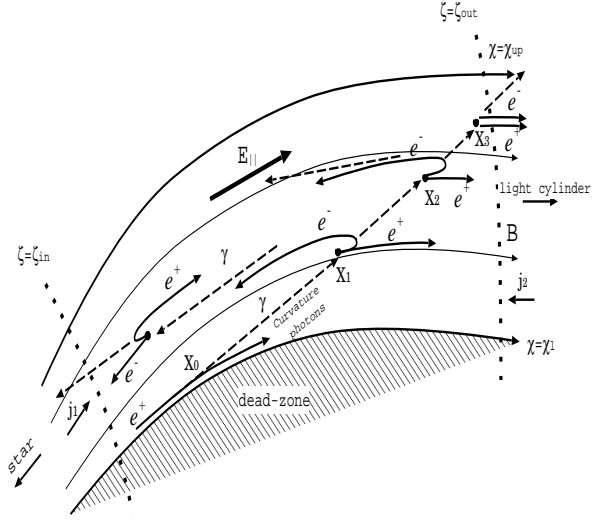}   
\caption{The pair creation cascade model in the meridional plane.    
The electrons or positrons are accelerated by the electric field along
the magnetic field lines and emit the $\gamma$-rays. 
 In the meridional plane, the $\gamma$-rays radiated by the
curvature process are beamed in the direction of local magnetic
field. The radiated $\gamma$-rays may convert into the pairs by the    
pair creation process. The new born pairs are also accelerated and
emit the $\gamma$-rays. The $\chi_{l}$, $\chi_{up}$, $\zeta_{in}$, and 
$\zeta_{out}$ present the lower, upper, inner, and outer boundaries of
the gap, respectively.}   
\label{pmodel}   
\end{center}   
\end{figure}   

To calculate the source function $S(\mathbf{r})$ and $\gamma$-ray
spectrum, we apply the simulation code developed 
in Paper2 (but we will revise the code for some effects as described
later): \textbf{Step1)} We calculate  number ($\equiv G_0^i$) of the emitted
photons per unit time per unit volume at position
$\mathbf{X}_0$ (see Fig.\ref{pmodel}). The $\gamma$-rays are emitted 
in the direction of the particle motion $\mathbf{v}$ at $\mathbf{X}_0$. 
\textbf{Step2)} We calculate photon  number 
converted into pairs in a distance  $s_0(\sim\varpi_{lc})$ by 
$\delta G^{i}=G_0^{i}[1-\exp(-\int_0^{s_0}1/l_pds')]$.  
\textbf{Step3)} We divide  $\delta G^{i}$ photons into 
$n$ flux elements, and pick up $n$  
points on the three-dimensional path length  
($s_i<s_0, i=1,\cdots n$)  following Monte Carlo method (\S\S\ref{pmean}).  
The pair creation point $\mathbf{X}_{i}$ in the meridional plane is  
deduced from $|\mathbf{X}_i-\mathbf{X}_0|=s_i\cos\theta_{\gamma}$,
where $\theta_{\gamma}=\cos^{-1}\sqrt{1-(\Omega\varpi/\varpi_{lc})^2}$. 
\textbf{Step4)} We  calculate the number of pairs produced 
 per unit time per unit volume as the source  term $S(\mathbf{X}_i)$.  

In this paper, we take into account the following effects:
\begin{itemize}
\item the energy spectrum  of the curvature photons emitted by the individual particles, 
\item the soft photon field proportional to the inverse square of the
distance (that is, inhomogeneous) and the soft photons propagate 
radially from the stellar surface (that is, anisotropic),  
\end{itemize}
which were ignored in Paper2 for simplicity. In the following subsections, we
detail how the above effects are built onto the cascade model.

\subsubsection{Curvature spectrum}
To give $G_0^i$ in \textbf{Step3)}, we define the following 
$\gamma$-ray photon number densities in the dimensionless energy interval 
between $x_i$ and $x_{i+1}$, 
\begin{equation} 
G_0^i=\frac{\int_{x_i}^{x_{i+1}}P_c(x)  dx}{(x_i+x_{i+1})/2} 
=\frac{\sqrt{3}e^2\Gamma}{hR_c}\frac{\int_{x_i}^{x_{i+1}}F(x)dx}{(x_i+x_{i+1})/2}
 ~~~(i=1,\cdots,n_1-1),
\end{equation} 
where $x_i=E_{\gamma}^{i}/E_c$. We define $x_1=0$ and 
select  each value of  $x_i$ $(i=2,\dots n_1)$ so as to satisfy 
$\int_{x_{i}}^{x_{i+1}}F(x)dx=a_1/n_1$, where
$a_1=\int_0^{\infty}F(x)dx=8\pi/3^{5/2}$. Instead of $x=\infty$,
 we define $x_{n_1+1}\equiv (3x_{n_1}-x_{n_1-1})/2$ and 
$G_{0}^{n_1}\equiv a_1/x_{n_1+1}$. For each  bin $G_0^{i}$, we carry out
the processes \textbf{Step1)}$\sim$\textbf{Step4)}. In this paper, we
adopt $n_1=20$ and $n=20$ in \textbf{Step3)}. 

\subsubsection{Pair creation cascade model}
\label{pmean}
The radial dependence in the
mean free path of equation (\ref{meanfree}) can be written as
$l_p\propto (1-\cos\theta_{X\gamma})^{-1}r^{2}$, where $r^{2}$ 
arises due to soft photon field diluting with the radial distance and 
$\theta_{X\gamma}$ of equation (\ref{coliang}) 
is the collision angle between an $X$-ray
photon and a $\gamma$-ray photon. 
Because $\cos\theta_{X\gamma}\sim \cos\theta_{em}$ is satisfied as long as the
propagating distance $s$ is smaller than the radial distance
$r\sim0.5\varpi_{lc}$, 
the radial dependence of the collision angle 
for  $\gamma$-rays emitted at the same
position is less important. In this paper, therefore, we adopt
$\theta_{X\gamma}=\theta_{em}$, 
which are collision angle at the emission point 
of the $\gamma$-rays considered. 
We take into account 
the fact that  the value of $\theta_{em}$  depends on the  emission point 
and the emitted direction (inwardly or outwardly) of the $\gamma$-rays. 
 
Using Monte Carlo method, we simulate the pair-creation
position in the gap. The probability that a $\gamma$-ray photon 
makes a pair between distance $s$ and $s+ds$ from the emission point
is  
\begin{equation}
P_{p}ds=\frac{\exp(-\int_0^s1/l_p(s')ds')}{l_p(s)}ds.
\label{posi}
\end{equation}
Using a function [$\equiv f(y)$], we transform 
 uniformly distributed  random numbers, 
$\{y_1,y_2, \dots, y_{n}\} \equiv\mathbf{y}$, into random numbers, 
$\{s_1,s_2,\cdots, s_{n}\} \equiv \mathbf{s}$, following the
probability $P_p$.  We obtain 
the function $f(y)$ by solving  the  equation (Press et al., 1988) 
\begin{equation} 
\left|\frac{dy}{ds}\right|=\frac{1}{l_p}\exp(-\int_0^s1/l_pds'). 
\label{press}
\end{equation} 
For the  present case, $l_p\propto r^2$, we can perform the
integration in equation (\ref{press}) analytically. As a result, we
obtain the  function $f(y)$ as  
\begin{equation} 
s=f(y)=r_0\sin\theta_{em} 
\tan\left[-r_0\frac{\sin\theta_{em}\mathrm{ln}y}{c_1}  
+\mathrm{atan}\left(\frac{1}{\tan\theta_{em}}\right)\right] 
-r_0\cos\theta_{em},
\label{conversion}  
\end{equation} 
where 
\[
c_1=(1-\cos\theta_{X\gamma})\frac{2\pi^2R^2_{eff}}{h^3c^3}
\int_{E_{th}}^{\infty}dE_{\gamma}\frac{E_{X}^2}{\exp(E_X/kT_s)-1}\sigma_p.
\]

The transformation represented by equation (\ref{conversion}) 
gives a random number $s$
of  $0<s<\infty$ with a uniformly selected random number
 $y$ of $0<y<1$. Because we are concerned with 
 the pair creation positions between $0<s<s_0\sim\varpi_{lc}$, 
we define uniformly
distributed random numbers as $\mathbf{y}^1\equiv(c_y-1)\mathbf{y}+1$,
where 
\[  
c_y=\exp\left\{-\frac{1}{r_0\sin\theta_{em}}\left[\mathrm{atan}  
\left(\frac{s_0+r_0\cos\theta_{em}}{r_0\sin\theta_{em}}  
\right)-\mathrm{atan}\left(\frac{1}{\tan\theta_{em}}\right)\right]  
\right\}.
\]  
A series of the transformations, $\mathbf{y}\Rightarrow
\mathbf{y}^1\Rightarrow \mathbf{s}=f(\mathbf{y}^1)$,  produces
the pair creation positions $\mathbf{s}$ following the probability
$P_p$  between $0<s<s_0\sim\varpi_{lc}$. 

Following above Monte Carlo method,  we choose $n$ pair-creation 
positions  for $G_0^i$ photons emitted at the position considered. 
On each pair-creation position, the pairs produced per 
unit time and per unit volume is given by 
$G_0^i[1-\exp(-\int_0^{s_0}1/l_pds')]/n$. 

\subsection{Boundary conditions}
\label{boundary}
 We introduce the conventional dimensionless variables:   
\begin{equation}  
\tilde{\mathbf{r}}=\frac{\omega_p}{c}\mathbf{r},  \ \ \ \  
\omega_p=\sqrt{\frac{4\pi e^2}{m_e}\frac{\Omega B_{lc}}{2\pi ce}},  
\end{equation}  
\begin{equation}  
\tilde{\Phi}_{nco}=\frac{e}{m_ec^2}\Phi_{nco}, \ \ \ \  
\tilde{\mathbf{E}}=-\tilde{\nabla}\tilde{\Phi}_{nco},  
\end{equation}  
\begin{equation}  
j_{\pm}=\frac{ev_{||\pm}N_{\pm}}{\Omega B/2\pi}, \ \ \ \
\tilde{\mathbf{B}}  
=\mathbf{B}/B_{lc},  
\end{equation}  
\begin{equation}  
\mathbf{\beta}=\frac{\mathbf{v}}{c},  
\end{equation}  
and  
\begin{equation}  
\tilde{\rho}=\frac{\rho}{\Omega B_{lc}/2\pi
c}=\frac{\tilde{B}}{\beta_{||}}  
(j_+-j_-), \ \ \ \ \tilde{\rho}_{GJ}=-\tilde{B}_z.
\label{lcharge}  
\end{equation}  
The Poisson equation (\ref{basic1}) and the continuity equation
(\ref{basic2}) are rewritten as 
\begin{equation}  
\tilde{\triangle}\tilde{\Phi}_{nco}=-[\frac{\tilde{B}}{\beta_{||}}(j_+-j_-)
 +\tilde{B}_z],  
\label{nbasic} 
\end{equation}  
and  
\begin{equation}  
\tilde{\mathbf{B}}\cdot\tilde{\nabla}j_{\pm}=\pm\tilde{S}(\tilde{\mathbf{r}}),
 \label{nbasic2}  
\end{equation}  
respectively.

In the meridional plane, the outer gap  has four boundaries
 called  as upper, lower, inner and outer
boundaries (see Fig.\ref{pmodel}). The upper and lower
boundaries are defined by the magnetic surfaces labeled by 
 $\chi_{up}$ and $\chi_{l}$, respectively.
 The inner $[\equiv \zeta_{in}(\chi)]$ and outer $[\equiv
\zeta_{out}(\chi)]$ boundaries are functions of $\chi$. 
For example, $\zeta_{out}(\chi)$=constant represents the
outer boundary defined by a curve line perpendicular to the magnetic
field lines. 

We impose  some conditions on the four boundaries. 
The inner and the outer boundaries are defined by the surfaces on
which the field-aligned electric field $\tilde{E}_{||}$ vanishes,
\begin{equation}
\tilde{E_{||}}(\zeta_{in})=\tilde{E_{||}}(\zeta_{out})=0.
\label{cond2}
\end{equation}
With $Con$3 in \S\ref{beq}, we have postulated that 
the condition $\tilde{\mathbf{E}}\cdot\tilde{\mathbf{B}}=0$ 
holds between the stellar surface and the outer gap. 
The uniform value of $\tilde{\Phi}_{nco}$ over the
stellar surface propagates to the inner, upper and lower
boundaries. By setting an arbitrary constant for $\tilde{\Phi}_{nco}$ on the 
stellar surface to be zero, we impose 
\begin{equation}  
\tilde{\Phi}_{nco}(\zeta_{in})=\tilde{\Phi}_{nco}(\chi_{up})=  
\tilde{\Phi}_{nco}(\chi_{l})=0.  
\label{cond1}  
\end{equation} 
As  discussed in \S\S\ref{inpart}, there is a possibility that the
condition $\tilde{\mathbf{E}}\cdot\tilde{\mathbf{B}}=0$ is not held between 
the stellar surface and the inner boundary.   For such a case, it is  difficult
to determine the boundary condition on the inner boundary of the outer
gap. However, because  the potential drop between the
stellar surface and the inner boundary is expected  much smaller than 
the potential drop of whole outer gap as discussed in
\S\S\ref{inpart}, the condition (\ref{cond1}), $\tilde{\Phi}_{nco}=0$,
at the inner boundary is a good treatment.

The continuity equation (\ref{nbasic2}) satisfies the conservation law of
the longitudinal current density, that is, the total current density 
$j_++j_- $($\equiv j_{tot}$) is constant along a field line.   
 This total current $j_{tot}$ can be  rewritten by sum of   external
 and internal components. The external component is separated
 into two parts:   
\begin{enumerate}  
\renewcommand{\theenumi}{(\arabic{enumi})}   
\item $j_1$, the current carried by the positrons (e.g. originating 
in sparking on the stellar surface) coming into the gap through the 
inner boundary,  
\item $j_2$,  the current carried by the electrons (e.g. originating in   
 pulsar wind region) coming into the gap through the outer boundary. 
\end{enumerate}  
In the following, we call $j_1$ and $j_2$  as the  external positron 
 and  electron components, respectively. 
The internal component ($\equiv j_g$) is carried by the electrons and
positrons produced in the gap. In terms of $(j_g,j_1,j_2)$, the total
current density becomes 
\begin{equation} 
j_{tot}(\chi)=j_1(\chi)+j_2(\chi)+j_g(\chi).
\end{equation}

The current should be determined by some global conditions, because 
the outer gap, the polar cap and the pulsar wind interact each
 other through the current. By parameterizing external positron and
electron components, 
we impose the conditions as follows 
\begin{equation}  
j_{+}(\zeta_{in})=j_1(\chi),  
\label{cond3}
\end{equation}  
and  
\begin{equation}  
j_-(\zeta_{out})=j_2(\chi). 
\label{cond4}
\end{equation}  

We have four model parameters, i.e. the inclination angle 
($\alpha_{inc}$) and the three current components $(j_1,j_2,j_g)$.
Because we have Dirichlet-type and Neumann-type conditions on the inner
boundary, we solve the position of the inner boundary. 
We fix the lower boundary to the last open line. In the present model,
 we obtain unique positions of the inner and outer
boundaries by giving the three current components and the position of the upper
boundary.  In the calculation, however, we give the position of the
outer boundary instead of the internal current ($j_g$) and solve 
the trans-field distribution of $j_g$,  because it is difficult to
manage the distribution of $j_g$, which is affected by the pair 
creation cascade, by hand.

We consider the following particular solution in the present paper. 
In general, the internal current $j_g$ increases as the outer boundary
shifts toward the light cylinder if one fixes the position of the upper
boundary. In the present electrodynamical model, 
the solutions satisfying the boundary conditions disappear if 
a large internal current $j_g$ flows on a magnetic field line. 
We regard that the gap having a marginal structure represents 
the state that actually appearers. This critical value of the 
internal current component is denoted by  $j_{crit}$. 
If one gives the position 
of the upper boundary ($\chi_{up}$) and  the external current
components ($j_1,j_2$), the trans-field distribution
of $j_g$ having  the critical internal components $j_{crit}$ 
and the positions of the inner ($\zeta_{in}$) and outer ($\zeta_{out}$)
boundaries are determined. 

As we just mentioned, the solutions satisfying the boundary conditions
disappear if the internal component $j_g$ becomes too large 
($>0.1\sim 0.2$), because the produced
pairs may make the super Goldreich-Julian charge density and may quench
the gap. On the other hand, we can significantly increase the external 
component $j_1$ or $j_2$ because the injected particles only shift 
the gap outward and inward. For example, if the external electron 
component $j_2$ increases, the effective charge neutral surface
 and  the outer gap position shift inside relative to the
conventional  null charge surface of the Goldreich-Julian charge density (see
\S\S\ref{result1}). 

A solution is given by a self-consist connection between 
the field-aligned  electric field  determined by the distribution 
of the charge density (i.e. the pair creation cascade), the pair
creation cascade
controlled  by the curvature radiation process 
(i.e. the Lorentz factor in the gap),
 and the Lorentz factor calculated from the electric field.

The expected $\gamma$-ray flux on the  Earth is given by
\begin{equation}
F(E_{\gamma})=\frac{1}{\triangle\Omega d^2}N_{\gamma}(E_{\gamma})
\label{expect}
\end{equation}
where $\triangle\Omega$ is solid angle of the $\gamma$-ray beam and
$N_{\gamma}(E_{\gamma})$ is the total photon number emitted  per unit energy
per unit time,
\begin{equation}
N_{\gamma}(E_{\gamma})
=\int_{V_{tot}}\frac{P_c(\mathbf{r}, E_{\gamma})
N_{\pm}(\mathbf{r})}{E_{\gamma}}dV,
\end{equation}
where $V_{tot}$ represents the total  volume of the emission region,
which includes  both  inside and
outside the gap.  In the present two-dimensional model, the
solid angle of the $\gamma$-ray beam is estimated as
 $\triangle\Omega\sim 2\pi D_{\perp}/r_{out}$,
where $D_{\perp}$ and  $r_{out}$ are, respectively, 
 the solved trans-field thickness and 
the radial distance to the outer boundary of the outer gap.
The total photon number  $N_{\gamma}(E_{\gamma})$ emitted in the gap
 depends on  the azimuthal spread angle
($\equiv\triangle\phi$) of the outer gap, which is unsolved in the
present model. Because there are no detail discussions for the
pair creation cascade process with the three dimensional geometry up to
now, it is  difficult to estimate the azimuthal spread angle of the gap
in the present two-dimensional model. 
Therefore, we use  the azimuthal spread
angle as a model parameter to be determined by the observed flux.
The observed emission phase in the
light curves (Kanback et al. 1994; Fierro et al. 1998)  and
the previous geometrical studies (Romani 1996; Cheng et
al. 2000) for the light curves model have predicted 
$\triangle\phi\sim \pi$~radian.

\section{Results}
\label{results} 
In this section, we present the results where we use  
 the parameters of the Vela pulsar whose $X$-ray
field at the outer gap is deduced from the observations. 
From $ROSAT$ (\"{O}gelman, Finley 
\& Zimmerman 1993), $RXTE$ (Harding et al. 2002)  and $Chandra$ (Pavlov et
al. 2001) observations of the Vela pulsar, the spectrum in 
$X$-ray band  is expressed by two components;
power-low component and surface black body component. 
Because the former component is much fainter than  the latter 
 component, we neglect the former contribution to the $X$-ray 
field of the pair creation process.  \"{O}gelman et al.(1993) fitted the
thermal component with  $T^{\infty}\sim 1.7$~MK and
$R_{eff}\sim 1.3 (d/0.25\mathrm{kpc})^2$~km, where $d$ is the distance 
to the Vela pulsar. In this paper, we
adopt $T^{\infty}=1.7$~MK and $d=0.25$~kpc (Cha, Semback \& Danksm 1999).

In the following subsections, we examine the dependence of the
$\gamma$-ray spectrum on the current components by fixing firstly the
critical internal component $j_{crit}$ (\S\S\ref{result1}) and secondly the
external components, $j_1$ and $j_2$ (\S\S\ref{result2}). In
\S\S\ref{diagnosis}, we compare the
results with the observed spectrum of the Vela pulsar, and diagnose the
outer gap geometry, the current and the inclination angle.

\subsection{Gap structures for  various external components}
\label{result1}
\begin{figure}
\begin{center}   
\includegraphics[width=10cm, height=8cm]{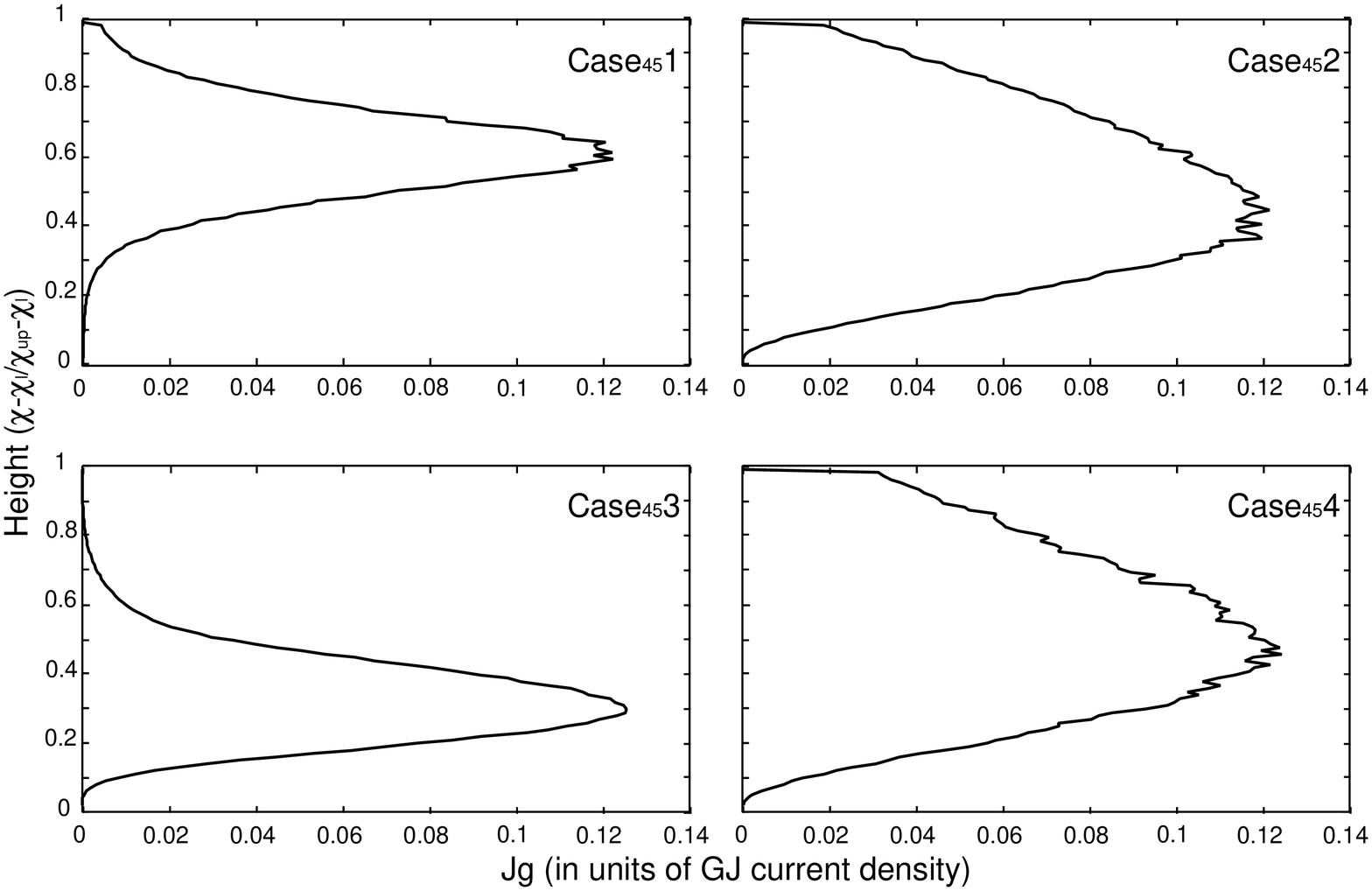}   
\caption{The trans-field structures of the internal component $j_g$ for
\case1$\sim$\case~4.}   
\label{current}   
\end{center}   
\end{figure}   

In this subsection we fix the inclination angle at
$\alpha_{inc}=45^{\circ}$ and the critical internal component at $j_{crit}\sim
0.12$. To examine the effects of the external components ($j_1,j_2$),
 we consider the following four cases:
\begin{enumerate}  
\renewcommand{\theenumi}{\case~\arabic{enumi}:}  
\item $(j_1,j_2)=(10^{-5},10^{-5})$ for  $\chi_{up}\ge \chi\ge
\chi_{l}$, 
\item $(j_1,j_2)=(0.05,0.05)$ for $\chi_{up}\ge \chi\ge \chi_{l}$,  
\item $(j_1,j_2)=(0.1,0.0)$ for $\chi_{up}\ge \chi\ge \chi_{l}$,  
\item $(j_1,j_2)=(0.0,0.1)$ for $\chi_{up}\ge \chi\ge \chi_{l}$.  
\end{enumerate}
In  \case~2$\sim$\case~4,  the external current is 10\% of the
Goldreich-Julian value, which is carried by  
both  components (for \case~2), the  positron component (for \case~3),
 and the electron component (for \case~4). 
In \case~1, we consider a case of very small  external component, 
that is, whole  current is carried by the internal component, 
$j_{tot}(\chi)\sim j_g(\chi)$. 

Fig.\ref{current} shows the trans-field distribution 
of the internal component $j_g$ for \case~1$\sim$\case~4. 
The abscissa and ordinate refer, respectively, the current density 
in units of the Goldreich-Julian value
 and the hight from the last open line (i.e. the lower boundary).  
The total current density on each magnetic field line is
represented by $j_{tot}(\chi)=j_g(\chi)+j_1+j_2$. So, for each case,   
the total current density  on the magnetic field line where the
internal component has the critical value ($j_{crit}\sim0.12$) is
 $j_{tot}\sim 0.12$ for \case~1 ($j_1=j_2=0$) and $j_{tot}\sim 0.22$ 
for \case~2$\sim$\case~4 ($j_1+j_2=0.1$). 
 
\subsubsection{General features}
\label{general}
\begin{figure}
\begin{center} 
\includegraphics[width=10cm, height=8cm]{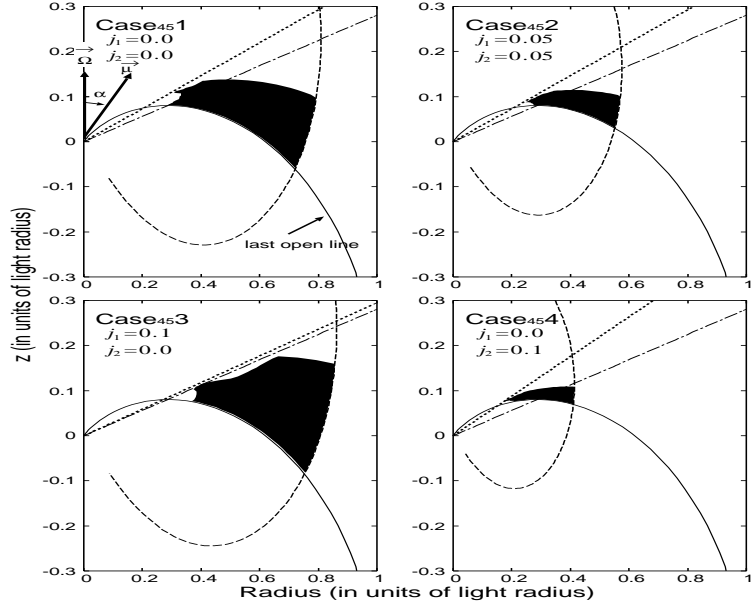} 
\caption{The geometries of the outer gap (filled region in each panel)
for \case~1$\sim$\case~4.  
The solid and dashed and dashed-dotted lines show the last open line, 
the curve line representing the outer boundary and the conventional 
null charge surface, respectively. 
On the  dotted line in each panel, the condition,  
$j_{crit}+j_2-j_1=\tilde{B}_z/\tilde{B}$, where $j_{crit}\sim0.12$, 
 is satisfied .} 
\label{nvagap} 
\end{center} 
\end{figure} 

\begin{figure}
\begin{center}   
\includegraphics[width=10cm, height=8cm]{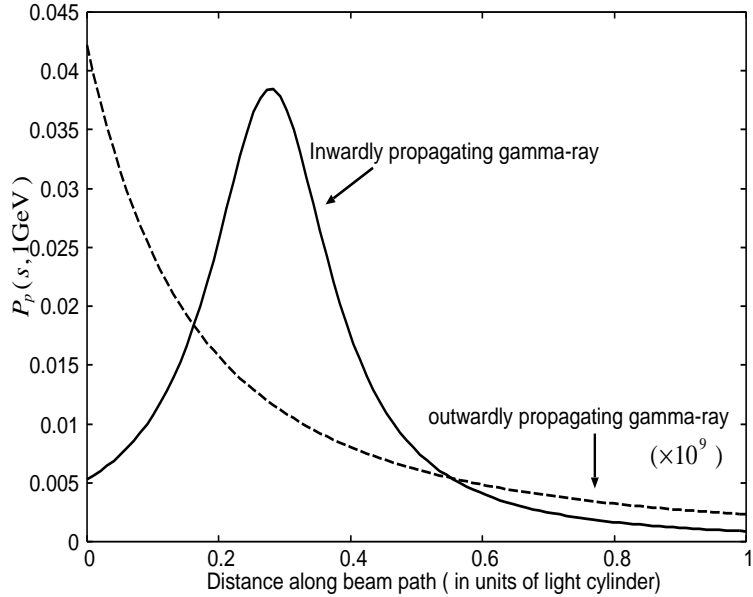}   
\caption{Par-creation probability of a $\gamma$-ray 
with 1GeV emitted from the null point on the last open line. 
The solid and dashed lines refer, respectively,  the probability 
of the inwardly and outwardly propagating $\gamma$-rays 
in the Vela magnetosphere. The
dashed line shows the value increased  $10^{9}$ times of the probability.}   
\label{mf1GeV}   
\end{center}   
\end{figure} 

\begin{figure}  
\begin{center}   
\rotatebox{270}{\includegraphics[width=8cm, height=10cm]{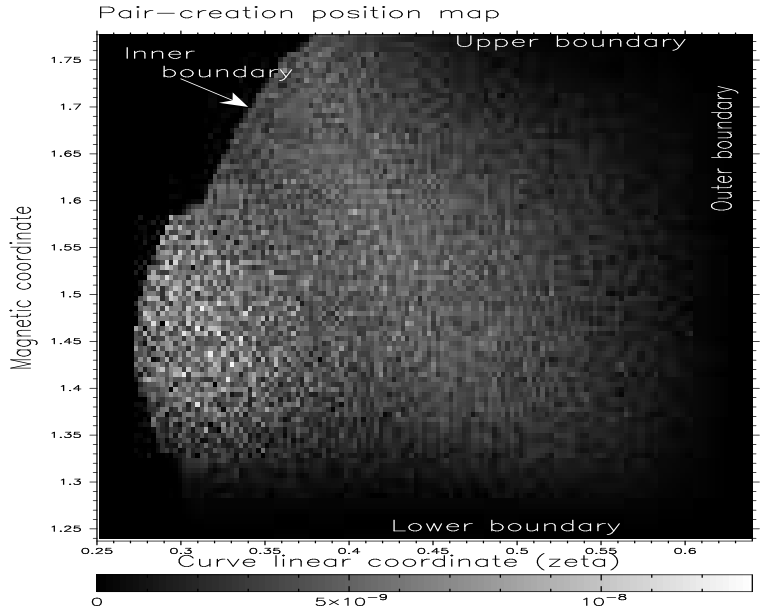}}   
\caption{The pair creation position map in the gap of \case~2. 
The whiteness refers the particles flux  (divided by $\Omega
B/2\pi e$) produced  per unit length [/cm] due to the pair creation 
at each position.}
\label{map}   
\end{center}   
\end{figure} 

The filled regions in Fig.\ref{nvagap} show the outer gap in the
magnetosphere for \case~1$\sim$\case~4. The inner boundary in each
panel  satisfies both  conditions $\tilde{E}_{||}=0$ and
$\tilde{\Phi}_{nco}=0$. The solid and  dashed-dotted lines 
in each panel are the
last open line and  the conventional null charge surface, respectively. 
 The  outer boundaries of \case~1,~2,~3, and 4 are  defined by the 
 curved lines (the dashed lines) labeled by $\zeta_{out}\sim0.9$, 
$\sim$0.65, $\sim$0.95,  and $\sim 0.45$, respectively.  

We  list the general features that were  already detailed   
in Paper2, in which $\alpha_{inc}$=0$^{\circ}$:  
\begin{itemize}
\item  The external component affects the gap position in the pulsar
magnetosphere.  As appeared in Fig.\ref{nvagap}, the typical gap position of \case~3 or
\case~4 is shifted outside or inside relative to the typical positions of
\case~1 and \case~2. Specifically, if $j_1=j_2$ such as \case~1 and
\case~2, the inner boundary of the lower
parts of the gap, where $j_g\sim0$, is located close to the conventional null charge
surface ($\rho_{GJ}=0$). On the other hand, if $j_1\neq j_2$ such as \case~3 or 
\case~4, the outer boundary shits outside or inside.  This is because
 the external component shifts the effective charge neutral surface 
 relative to the conventional null  surface.

\item As the  dotted lines,  on which $j_{crit}+j_2-j_1=\tilde{B}_z/
\tilde{B}$ with 
$j_{crit}\sim0.12$ is satisfied, shows, the cusp of the inner boundary 
of each case is located at  the  position where  the space charge density  
caused by the current is equal to the local Goldreich-Julian charge
density. The inner boundary shifts toward the stellar surface as the 
critical and/or  external electron components  increases.

\item As Fig.\ref{current} shows, the present model predicts that 
more  large current runs though the  middle part of the gap rather 
than the upper part predicted in the CHR86 arguments. 
The internal component $j_g$  increases with hight in the lower half
part of the gap, and decreases in the upper half part.  
This decrease is because most $\gamma$-rays emitted in the lower 
part of the gap  escape from the inner boundary  or the outer boundary
before reaching the upper part. Another reason is 
that the field-aligned  electric field, the Lorentz
factor and in turn the emitted photon energy decrease with 
 increasing  hight in the upper half part.
\end{itemize}

\subsubsection{Gap geometry \& pair creation mean-free path}
\label{gapgeo}
In Paper2,  the trans-field thickness of the outer gap was not
sensitive to the external component (for example fig.6 in Paper2),  
because  we assumed a homogeneous $X$-ray field.
On the other hand,  we find in Fig.\ref{nvagap} that the trans-field 
thicknesses
of \case~2 and \case~4 are much smaller than those of \case~1 and
\case~3. This difference in the thickness is 
 explained in terms of  the  mean-free path depending on the radial distance   
and the emission direction, which were  ignored in Paper2, as follows. 

Fig.\ref{mf1GeV} shows how the pair-creation probability of equation 
(\ref{posi})  develops along the path of 1GeV $\gamma$-rays emitted 
from  the null  
point on the last open line. The solid and dashed lines refer the
probabilities of a photon emitted inward and outward, respectively. 
The dashed line shows the value increased  $10^{9}$ times of 
 the probability. We find that the mean free path of the outwardly  
propagating $\gamma$-rays is much longer
 than that of the inwardly propagating ones, in which  a photon  out
of about $10^3$ photons   converts into a pair after
 running about the light radius. 
Fig.\ref{map} shows the pair creation position map of \case~2
($j_1=j_2=0.05$), where  we do not plot  the pairs created  
outside the gap.  In Fig.\ref{map}, the increase in the whiteness refers
the pair flux produced per unit length at each position.  
As we  expected,  the inwardly propagating $\gamma$-rays make many 
 pairs  around  
the inner boundary and the middle part of the gap. 

The large difference in the mean-free path between the outwardly and
inwardly propagating $\gamma$-rays  is mainly attributed to
 difference in the  collision angle $\theta_{X\gamma}$ with the
surface X-ray photon.  
For example, the collision angle  of the  $\gamma$-ray photon 
emitted from the conventional null point on the 
last open line of  $\alpha_{inc}=45$$^{\circ}$ becomes
$\theta_{X\gamma}\sim 22.8$$^{\circ}$ for the  outwardly  propagating
and becomes $\theta_{X\gamma}\sim 157.2$$^{\circ}$ for the
 inwardly propagating  $\gamma$-rays. So, the pair creation
processes occur  with  tail-on like 
collision  ($1-\cos\theta_{X\gamma}\sim 0$) 
for the outwardly and head-on like collision 
($1-\cos\theta_{X\gamma}\sim 2$) for the inwardly propagating
$\gamma$-rays. 
 For the outwardly propagating $\gamma$-rays with 1GeV,  
the threshold energy of the $X$-ray photon of the pair creation  is given 
by $E_{th}=2(m_ec^2)^2/(1-\cos\theta_{X\gamma})/\mathrm{1GeV}\sim6.7$keV,
which indicates the deep Wien region  of the Planck distribution for the
Vela pulsar ($kT_{s}\sim0.15$keV). For the inwardly propagating
$\gamma$-rays, the threshold energy becomes $E_{th}\sim 0.27$keV. 
This difference in the threshold energy and  the resultant difference
in the $X$-ray photon number for the collision mainly produce 
the large difference in the mean free path.  
 
Because a large production of the  pairs immediately quenches  
the gap, the mean-free path of the pair creation controls the
gap size. As a results,  \case~2 and \case~4, in which many  $\gamma$-rays 
are emitted inward by  the external electrons, 
 have much smaller gap thickness than \case~3, in which  many 
$\gamma$-rays  are emitted outward by  the external positrons.

\subsubsection{Electric structure}
\label{elest}
\begin{figure}
\begin{center}  
\includegraphics[width=10cm, height=8cm]{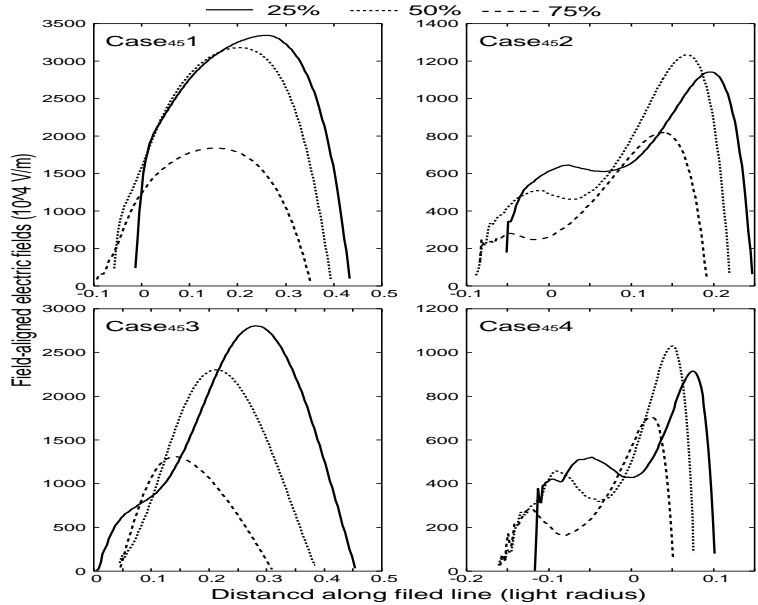}  
\caption{The electric field structure for \case~1$\sim$\case~4.   
The solid, dotted, and dashed lines in each panel show the distribution of  
the field-aligned electric field on the three different magnetic field lines, 
which locate, respectively, 25, 50, and 75\% of the gap trans-field thickness  
 measured from the last open field line. The abscissa represent the arc length
  with the origin at the null surface.}  
\label{electric}  
\end{center}  
\end{figure}  

Fig.\ref{electric} shows the distribution of the field-aligned 
electric field along the magnetic field lines for \case~1$\sim$\case~4. 
The solid, dotted and dashed lines
 represent, respectively, the field-aligned electric  field 
on the magnetic field through  the gap at 25, 50 and 75\% of the
trans-field thickness from the lower boundary. The abscissa refers 
the arc length from the conventional null charge surface, 
where the positive and negative values indicate outside and inside,
respectively, with respect to the null surface. 

Combining Fig.\ref{nvagap} and  Fig.\ref{electric}, we find a 
tendency  that the field-aligned electric field becomes strong 
with increasing
 trans-field thickness. For example, in \case~2, \case~3 and \case~4,
in which $j_{tot}\sim0.2$, \case~3 has both the largest trans-field
thickness and the strongest field-aligned electric field, on the other hand, 
 \case~4 has both smallest trans-field thickness and the weakest
field-aligned electric field.   The available 
potential drop ($\equiv\delta\Phi_{a}$ ) in the outer gap may be written
  as $\delta\Phi_{a}\sim \Phi_{a}\delta\theta_{gap}/\theta_p$, 
where $\Phi_{a}$ is the available potential drop on the stellar
 surface, $\delta\theta_{gap}$ is
the angle of the foot points of the open field lines through 
 the gap, and $\theta_{p}$ is the angle of the polar cap.
Because the angle $\delta\theta_{gap}$ increases with  
the trans-field thickness, the potential drop increases with the
trans-field thickness. As shown  in Paper2, the strength of the field-aligned 
electric field does not depend on the gap width along the field lines 
very much, because the effective charge density
$(j_+-j_-)\tilde{B}/\beta_{||}+\tilde{B}_z$ and the transverse term 
$-\tilde{\triangle}_{\perp}\tilde{\Phi}_{nco}$  in the Poisson
equation almost balance out in the gap (\S\S\ref{basicpoi}). 
As a results, the strength of the electric field tends to increase with 
the trans-field thickness. This effect is
 originally pointed out by the vacuum gap model of CHR86. 
 
From  Fig.\ref{nvagap} and  Fig.\ref{electric}, we also find that 
the field-aligned electric field of \case~3 is weaker than that of \case~1, although 
the trans-field thickness of \case~3 is larger. 
 This reflects the screening effects of the particles 
on the field-aligned electric field. The dimensionless 
effective charge densities  near the outer boundary for \case~1 and
\case~3 are  written as
$\tilde{\rho}_{eff}=\tilde{B}j_g/\beta_{||}-\tilde{B}_z$ 
and $\tilde{\rho}_{eff}=\tilde{B}(j_g+j_1)/\beta_{||}-\tilde{B}_z$, respectively. 
For \case~3,  the external positrons 
assist  in the screening near the outer boundary. 

As we have seen,  the combination of the effects of the 
 gap thickness and the screening  of the particles
 determine the strength of the field-aligned electric field in the gap. 

\subsubsection{$\gamma$-ray spectrum}
\label{gammaray}
\begin{figure}
\begin{center}  
\includegraphics[width=10cm, height=8cm]{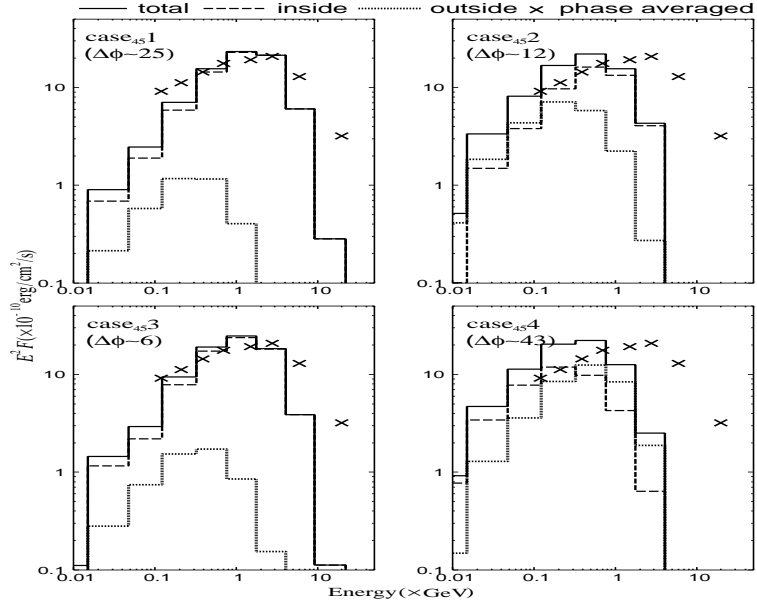}  
\caption{Calculated $\gamma$-ray spectra of the curvature radiation 
 for \case~1$\sim$\case~4 of the Vela pulsars. The solid line show the
expected spectrum, which are  composed of the  inside (dashed line) 
and outside (dotted line) emissions. The crosses show
the $EGRET$  phase-averaged spectrum.  The parameter 
 $\triangle\phi$ adjusts the calculated flux to
the observation. }  
\label{spectra}  
\end{center}  
\end{figure}  

 As we see in  Fig.\ref{map}, because most pairs are created near the
inner boundary, the emission region for the inwardly $\gamma$-rays in
the gap is restricted around the inner boundary, while the emission
region for the outwardly  $\gamma$-rays is whole outer gap. 
Furthermore,  we expect that almost inwardly
propagating $\gamma$-rays are converted into pairs by the magnetic or 
photon-photon pair creations when they pass close to the star. 
Therefore, we assume that  the contribution of the inwardly
$\gamma$-rays to the spectrum on the Earth is ignorable. 

Fig.\ref{spectra} shows the calculated  curvature spectra
of the outwardly propagating $\gamma$-rays for \case~1$\sim$\case~4.
 The solid lines in Fig.\ref{spectra} represent the expected
spectrum, which are composed of  the gap  emissions  (dashed line) and
the emissions (dotted line) from outside  the gap.  
The crosses in  Fig.\ref{spectra} show 
the $EGRET$  phase-averaged spectrum of the Vela pulsar. The adjusted
azimuthal spread angle to the observed flux for each case is
$\triangle\phi\sim25$~radian (\case~1), $\sim12$~radian (\case~2),
$\sim6$~radian (\case~3) and $\sim43$~radian (\case~4).

From Fig.\ref{spectra}, we find that the outside emissions (dotted 
lines) of \case~1 and \case~3 are fainter  than the inside
 emissions (dashed lines).  As the  outer gap geometries 
in Fig.\ref{nvagap} show, the outer boundary of 
\case~1 or \case~3 is located near the light cylinder. 
Therefore, the outside emission region is  smaller than the outer gap.  
 For \case~2 and \case~4, on the other hand,  
the outside emissions  contribute to   the total spectrum 
below 1GeV, because the size of the  
outside emission region is similar to or larger than  the gap size. 

Combining  Fig.\ref{electric} and  Fig.\ref{spectra}, we see that the
spectral hardness  reflects directly the strength of 
the field-aligned electric field.  
For example, the strongest (or weakest) electric field of \case~1 (or
\case~4) in all cases produces the hardest (or softest) spectrum. 
We find that the spectrum becomes to be soft as the currents
increase. Especially, by comparing the spectra of \case~1 and
\case~4, we see that the spectrum sharply softens as the electron
injection at the outer  boundary increases. 
 By comparing the spectra of  \case~1 and \case~3, 
on the other hand, we also find that the increasing in the positronic 
component gradually softens the spectrum.

\subsection{Dependence on the internal current $j_g$}
\label{result2}
\begin{figure} 
\begin{center}    
\includegraphics[width=10cm, height=8cm]{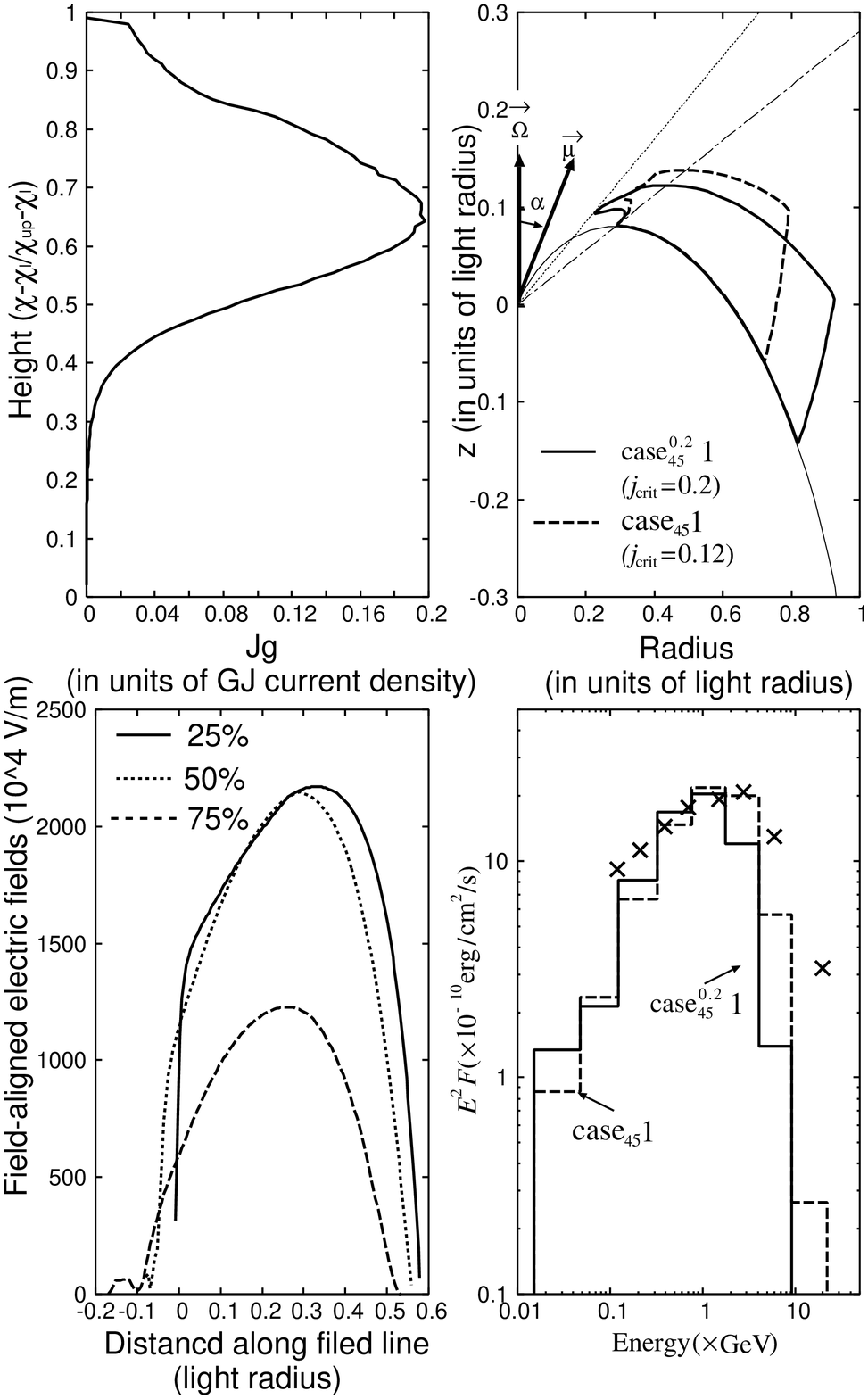}    
\caption{The solved outer gap ($\mathrm{case}_{45}^{0.2}~1$) 
of $(j_1,j_2)=(0,0)$ and $j_{crit}\sim0.2$. The solid line in 
each panel shows the trans-field distribution of the internal current
(upper left), the gap geometry (upper right), distribution of the 
field-aligned electric field (lower left) and the total spectrum (lower
right). For comparison, we also display the gap geometry and the total
spectrum of \case~1. }    
\label{gapcom}    
\end{center}    
\end{figure}    
In the last section, we have investigated the dependence of the
spectral hardness on the external components $(j_1, j_2)$. 
To clear the dependence of the spectrum on the critical internal 
component $j_{crit}$, we also calculate 
the  case that $j_1=j_2=10^{-5}$ and $j_{crit}\sim0.2$ (denoted as
$\mathrm{case}_{45}^{0.2}~1$) and compare the results with that of
\case~1 ($j_1=j_2=10^{=5}, j_{crit}\sim0.12$). The results of 
$\mathrm{case}_{45}^{0.2}~1$ are  summarized in Fig.\ref{gapcom}. 
The upper left, upper right, lower left and lower right panels show 
the trans-field distribution of the internal component, the gap geometry,
the distribution of the field-aligned electric field,
 and the total spectrum, respectively. 
For comparison, we also display the gap geometry and the total
spectrum of \case~1 with  dashed lines in Fig.\ref{gapcom}.
 
As Fig.\ref{gapcom} shows, the larger critical internal component,
that is $\mathrm{case}_{45}^{0.2}~1$, is predicted by a gap having 
the smaller trans-field thickness and longer width along the magnetic 
field lines. We  expect that the gap  of $\mathrm{case}_{45}^{0.2}~1$ has   
a longer  gap width $W_{||}$ than that of \case~1, because
 the internal component $j_g$ increases with the  gap width. 
For \case~1, we adjust the positions of the outer 
and upper boundaries  so as to reproduce a marginal gap structure having
$j_{crit}\sim0.12$. The field-aligned electric field changes 
it's sign in the gap due to the screening effect of the particles 
if we shift the outer boundary outward relative to that of \case~1 
without changing the upper boundary. As Fig.\ref{gapcom} shows,
therefore, the gap of $\mathrm{case}_{45}^{0.2}~1$  has a smaller 
thickness and a longer width than those of \case~1.   

The decrease of the gap thickness causes a large transverse term 
$\triangle_{\perp}\Phi_{nco}\sim -\Phi_{nco}/D_{\perp}^2$.
 As discussed in Paper2, the transverse term reduces the screening 
effects of the electrons, because it acts as a positive charge 
in the Poisson equation. The internal component $j_g$ can increase 
until the transverse term marginally sustain the negative charged 
super-GJ region, where $|\rho|>|\rho_{GJ}|$, in the gap. 
The choice of a large critical internal component tends to 
result in a smaller thickness.  

Because the smaller trans-field thickness gives the smaller 
field-aligned electric field and furthermore because the larger 
internal component $j_g$  gives the stronger screening effects 
for the field-aligned electric field, the resultant $\gamma$-ray 
spectrum of $\mathrm{case}_{45}^{0.2}~1$ is softer than that of \case~1. 

We summarize the  discussions in \S\S\ref{result1} and \S\S\ref{result2}. 
The spectrum becomes soft as the current components, 
($j_{crit}, j_1, j_2$), increase due to the screening effect on the 
field-aligned electric field. 
Especially, the increase in the electron component $j_2$ at the outer 
boundary sharply softens
the spectrum.  On the other hand, the increase in the  positronic
component $j_1$  at the inner boundary gradually softens the spectrum. 
By increasing the positronic component $j_1$, therefore,  we can obtain 
a brighter emission with similar spectral 
hardness (see \S\S\S\ref{fluxext}). This dependence of the spectral
feature on the current allow us to diagnose what current flows  
in the gap of the observed $\gamma$-ray pulsars.

\subsection{Application to  the Vela pulsar}
\label{diagnosis}
In the present two-dimensional model, the gap width
 and thickness are linked so that we can discuss the gap 
structure in terms of the spectral characteristic; 
i.e. the cut-off energy and the flux. 
 By comparing the model spectra with the observed spectrum, we can diagnose
 the free model parameters, that is, the current
components $(j_{crit},j_1,j_2)$ and the inclination angle $\alpha_{inc}$.  
In this paper, we apply the theory to the Vela pulsar.  
 
In \S\S\S\ref{cutoff}, we study that the
$EGRET$ spectrum of the Vela pulsar predicts the cases that have 
only a few  electron  injections ($j_2\sim0$) at the outer boundary. 
 In \S\S\S\ref{fluxext}, we discuss  parameter spaces of
$\alpha_{inc}$ and  $j_1$, which explain the $EGRET$ spectrum, 
with $j_{crit}\sim0.1$, which is moderate value in the present
electrodynamical model.

\subsubsection{Flow pattern of the external current}
\label{cutoff}
From  Fig.\ref{spectra}, we find that the spectrum of \case~2, in
which $j_1=j_2=0.05$, $j_{crit}\sim0.12$ and
$\alpha_{inc}=45^{\circ}$, becomes to be very soft as compared with 
the observation; the spectral cut-off energy is $\sim0.5$GeV for \case~2 
and  $\sim 3$GeV for the observation.   We also find that the 
adjusted  spread angle for \case~2, $\triangle\phi\sim12$~radian, is
unrealistic. Starting from  these facts, 
we can rule out any outer gap having a large particle 
injection  at the outer boundary for the Vela pulsar, as follows.

Firstly, we do not expect the cases having  a larger current
components than
that of \case~2, because the spectrum softens as the current
increases as discussed in \S\S\ref{result1} and \S\S\ref{result2}. 

Secondly, we do not expect the cases that the electronic 
component $j_2$ is the same with that of \case~2, but  the positronic 
component  $j_1$ and/or the internal component $j_{crit}$ are  
smaller than those of \case~2.  This is because the 
adjusted azimuthal angle  $\triangle\phi$ 
increases with the decreasing $j_1$ and/or $j_{crit}$
 due to decrease of the outwardly migrating particles. Therefore, the 
unreality on the spread angle such as $\triangle\phi=12$~radian of 
\case~2  grows with decreasing  $j_1$ and/or $j_{crit}$.  

Finally, we can show that the hardness of the spectrum does not
depend on the inclination angle very much.
 The increase in the inclination angle shifts the null point on 
the last open line toward the stellar surface. This shift increases 
the strength of the magnetic field in the outer gap, 
and also  decreases the gap thickness with the mean-free path 
because the $X$-ray field becomes dense and because the collision of 
the inwardly propagating $\gamma$-rays approaches the head-on. 
The former effect tends to increase the strength of the 
field-aligned electric field,  the latter to decrease it, 
the two effects compensate each other. As a results, 
 the spectral hardness does not depend on the inclination 
angle very much. In fact, there is a tendency that the spectral
hardness softens with the inclination angle for the large electron
component at the outer boundary. Therefore, we can rule out 
any inclination angles with $j_2\geq0.05$ for the Vela pulsar.

As a result of the above three arguments, we can rule out any large 
electronic injection cases, $j_2\geq0.05$, and  we suggest 
 that only a few electrons ($j_2\sim0.0$) are injected into the 
gap at the outer boundary for the Vela pulsar.  

For $j_2\sim 0$, the decreasing of  the gap thickness with 
increasing the inclination angle is more moderate than the case of 
the large external electron case, because the mean free path of the
outwardly $\gamma$-rays is very longer than the gap thickness.
As a results,  we can see that the strength of the  
field-aligned electric field and resultant
 hardness of the spectrum tend to increase 
with the inclination angle for $j_2\sim0$.
 
\subsubsection{Diagnosis the current and the inclination angle}
\label{fluxext}
\begin{figure}
\begin{center}     
\includegraphics[width=10cm, height=8cm]{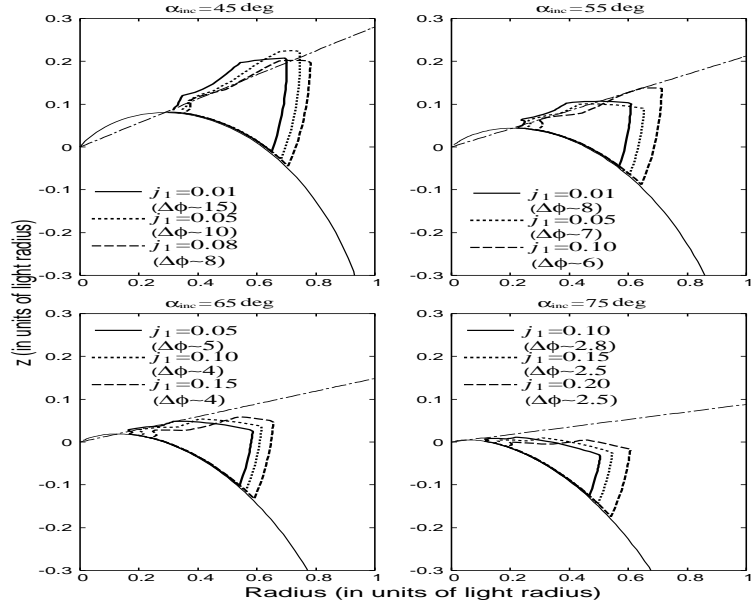}     
\caption{The calculated outer gap geometry in wide parameter  
space 45$^{\circ}$$<\alpha_{inc}<$75$^{\circ}$. We use $j_2=0$ and
$j_{cirt}\sim0.1$. The thin solid and dotted-dashed lines show the last-open
line and the conventional null charge surface, respectively.}     
\label{allgap}     
\end{center}     
\end{figure}  

\begin{figure}    
\begin{center}     
\includegraphics[width=10cm, height=8cm]{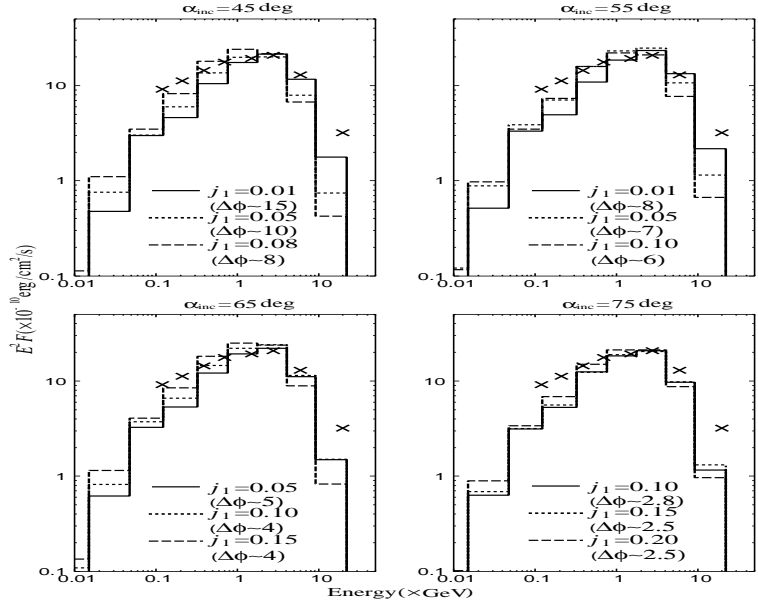}     
\caption{The calculated total  spectra with various inclination angles
and positron components. The external electron and  critical
internal components are $j_e=0$ and $j_{crit}\sim0.1$.  
 The calculated spectral hardness is explained with $j_1\leq0.08$
for $\alpha_{inc}=45^{\circ}$, $\leq0.10$ for
$\alpha_{inc}=55^{\circ}$, $\leq0.15$ for
$\alpha_{inc}=65^{\circ}$ and $\leq0.20$ for
$\alpha_{inc}=75^{\circ}$.}     
\label{allspect}     
\end{center}     
\end{figure}   

Following the  discussion in \S\S\S\ref{cutoff}, we fix the  
external electron component  $j_2$ at zero for the Vela pulsar. 
For $j_2\sim0$ with the present model, the solutions disappear if the critical 
internal component $j_{crit}$ exceed several ten percent  
of the Goldreich-Julian value. In this subsection, therefore,  
we adopt  $j_{crit}=0.1$ as a moderate value.  With $j_{crit}=0.1$ and
$j_2=0$, we diagnose other two model parameters, that is, the
 external positron  $j_1$ and the inclination angle $\alpha_{inc}$.

 Fig.\ref{allgap} and Fig.\ref{allspect} shows 
 the gap geometries and the spectra, respectively, for various 
positronic components and the inclination angles; 
upper left panel summarizes the results of  $j_1=0.01$ (solid), 0.05
(dashed) and 0.08 (dashed) with $\alpha_{inc}=45$$^{\circ}$, upper right panel $j_1=0.01$ (solid), 
0.05 (dotted) and 0.1 (dashed) with  $\alpha_{inc}=55$$^{\circ}$, lower left 
panel $j_1=0.05$ (solid), 0.10  (dotted) and 0.15 (dashed) with
$\alpha=65$$^{\circ}$  and lower right panel $j_1=0.10$ (solid), 0.15 (dotted)
and 0.20 (dashed) with $\alpha_{inc}=75$$^{\circ}$. 
 By comparing the observed cut-off energy, we can restrict the
positronic component $j_1$ such as $j_1\leq 0.08$  
for $\alpha_{inc}=45$$^{\circ}$, $\leq0.1$ for 55$^{\circ}$, 
$\leq0.15$ for 65$^{\circ}$, and $\leq0.20$ for 75$^{\circ}$.  
The result of the marginal case  for each inclination angle is 
 shown  with the dashed line. The values of the azimuthal spread angle 
$\triangle\phi$ of the gap written 
in Fig.\ref{allgap} and Fig.\ref{allspect} are adjusted to the observed flux.  

We notice that the increase in the inclination angle 
permits  larger  positronic component $j_1$.  
For only a few electronic injections at the outer boundary $j_2\sim0$,
the field-aligned electric field in the gap tends to become strong with
increasing the inclination angle as  described 
in \S\S\S\ref{cutoff}. Stronger field-aligned  
electric field makes us  possible to have larger positronic component 
$j_1$, because we can increase $j_1$ until 
the particles  screen  the field-aligned electric field 
so that the calculated spectral cut-off energy  
does not explain the observation.

For each positron external component $j_1$ of 
$\alpha_{inc}=45^{\circ}$ and $\alpha_{inc}=55^{\circ}$ in
Fig.\ref{allgap}, the adjusted  azimuthal angles $\triangle\phi$ are
near or lager than $2\pi$~radian which are unrealistic. 
Therefore, we do not expect  any cases of the critical internal
component of $j_{crit}\sim0.1$, 
which has been assumed in Fig.\ref{allgap}, for $\alpha_{inc}=45^{\circ}$ and
$\alpha_{inc}=55^{\circ}$. For smaller critical internal component
than $j_1\sim0.1$, the unreality of the azimuthal angle increases.  
 For more large critical internal component such as $j_{crit}\sim0.2$ of 
$\mathrm{case}_{45}^{0.2}~1$ in \S\S\ref{result2}, the calculated 
spectra become to be soft as compared with  the observation. On these 
ground, therefore, we safely rule out any cases of $\alpha_{inc} \leq
55^{\circ}$ for the Vela pulsar. 

For $\alpha_{inc}=65$$^{\circ}$, 
The adjusted azimuthal angle $\triangle\phi\sim 4$~radian of the marginal case
(dashed line) is slightly larger than $\triangle\phi\sim \pi$. Because 
the three-dimensional gap geometry  affects
the solid angle of the $\gamma$-ray beam and the expected flux on the Earth, 
we could deem the cases of $\alpha_{inc}=65$$^{\circ}$ acceptable 
for the Vela pulsar in the present 
two-dimensional model. We expect that the large positron component such as
$j_1\sim0.1$ runs through in the gap. 

For $\alpha_{inc}=75^{\circ}$,  
the adjusted  azimuthal angle of $\triangle\phi\sim2.5$~radian for $j_1=0.2$
explains the observation very well. Furthermore, for $j_1=0.1$  (the
solid line),  the expected  angle $\triangle\phi\sim 2.8$~radian 
 is also consistent with the
observation. Therefore, we find that both the observed spectral 
cut-off energy and flux  are explained by  $j_1\leq 0.2$ 
for $\alpha_{inc}=75$$^{\circ}$.  As we have seen,
 the upper limit  $j_1=0.2$ for  $\alpha_{inc}=75$$^{\circ}$ is determined 
by the spectral cut-off energy. On the other hand, 
the lower limit of the external positron $j_1$ is determined 
by a comparison between the model and the observed flux. 
Because the expected flux depends on the three-dimensional geometry, 
we should use a
three-dimensional model to give  the lower limit of $j_1$ for $\alpha_{inc}=75^{\circ}$.

In summary, we have shown by solving the two-dimensional gap structure
 that the strength of the field-aligned electric field and the
hardness of the produced $\gamma$-ray spectrum are strongly affected 
by the current through the gap.  By comparing the model spectrum with 
the $EGRET$ observation, we have successfully constrained 
the model parameters (the current and
the inclination angle) for the Vela pulsar, as follows.
Only a few external electrons (i.e. $j_2\sim0$) injected into the gap at the
outer boundary are  predicted by the $EGRET$ observation. 
We also safely rule out any cases of  $\alpha_{inc}\leq 55$$^{\circ}$. 
With   uncertainty due to the three dimensional effects, we conclude
that the Vela pulsar has the inclination angle $\alpha_{inc}$ 
greater than $65^{\circ}$. With
the moderate critical internal component $j_{crit}\sim0.1$, the positronic component
$j_1$ is restricted such as  $j_1\sim0.1$ for $\alpha_{inc}=65^{\circ}$ and 
$j_1\leq0.2$ for $\alpha_{inc}=75^{\circ}$. 

We can also explain the observed spectrum with greater inclination 
angles. Following the RY95 arguments, 
the inclination angle will be  not close to 90$^{\circ}$, 
because the radio pulsation of the Vela pulsar shows a single peak.  

\section{Discussion}
\subsection{Other high-energy emission processes}
In this section, we show the validity of the assumption 
 that the synchrotron and the inverse-Compton process
 are less important for the gap electrodynamics, and 
 calculate the inverse-Compton spectrum to compare the result
 with the recent observation in TeV range.
   
\label{emipro}
\subsubsection{Synchrotron process}

Pairs created  by the pair creation process in the gap 
 have significant pitch angle $\chi$ and  will radiate  the
synchrotron photons. Initial energy of the pairs is several GeV,
which is typically energy of the $\gamma$-ray photons radiated by the
curvature process of the accelerated particles. After the birth, the
pairs are accelerated to $\Gamma\sim10^{7.5}$ from the initial Lorentz
factor of $\Gamma_0\sim10^{3.5}$ and will have pitch angle of
$\chi\sim\Gamma_0
/\Gamma$. It follows that the ratio between the
 synchrotron power and the curvature power becomes
\begin{equation}
\frac{P_{syn}}{P_{curv}}=\left(\frac{eBR_c\sin\chi}{\Gamma
m_ec^2}\right)^2\sim10^{-3}\left(\frac{\Gamma_0}{10^{3.5}}\right)^2
\left(\frac{\Gamma}{10^{7.5}}\right)^{-4}\left(\frac{B}{10^5\mathrm{G}}
\right)^2\left(\frac{\Omega}{100\mathrm{s^{-1}}}\right)^{-2}
\left(\frac{R_c}{0.5R_{lc}}\right)^2.
\end{equation}
This  indicates that  the synchrotron process is less important for
the gap electrodynamics. However,  it is not denied
an another solution in which the synchrotron  radiation play
 an important role. For example, as we have seen, for the large current
case, the inner boundary is shifted toward stellar surface, where the
synchrotron process would be important because the  magnetic field
strength significantly increases. The present model
have predicted that the inner boundary of
the gap that is favored for the Vela spectrum does not locate near the
stellar surface (Fig.\ref{allgap}).

The large number of pairs will be  produced outside the gap. These pars radiate 
synchrotron photons, the typical energy of which is 
\begin{equation}
E_{syn}=\frac{3eh\Gamma^2B}{4\pi m_e c}\sin\chi=1.8\times
10^4\left(\frac{\Gamma_0}{10^{3.5}}\right)^2\left(\frac{B}{10^5\mathrm{G}}
\right)\sin\chi ~\mathrm{eV}.
\end{equation}
 It is, therefore, entirely justified to neglect the synchrotron
component emitted outside the gap when we considerer the
$\gamma$-ray spectrum above 100~MeV.

\subsubsection{Inverse-Compton process}
TeV photons are emitted by the inverse Compton (IC) scatterings of
infrared (IR) photons off the relativistic electrons and positrons, which emit 
GeV photons via the curvature process. For the
Vela pulsar,  Romani (1996) predicted the pulsed TeV flux at 
about  1\% of the pulsed GeV flux, which has been ruled out by Air Cherenkov
telescopes (Konopelko et al. 2005). 
Therefore, the theoretical prediction of TeV flux is important for
checking the  validity of the model.

We assume the IR field is isotropic and homogeneous, 
because the IR photons will be radiated by the synchrotron process 
of the pairs with a larger pith
angle. For isotropic case,  the relativistic particle upscatters the soft
photons to produce the following spectrum (Blumenthal \& Gould 1970)
\begin{eqnarray}
\frac{dN}{dtdE_{\gamma}}&=&\frac{3}{4}\sigma_T\frac{c}{\Gamma^2}
\frac{dN_{IR}}{dE_{IR}}\frac{dE_{IR}}{E_{IR}} \nonumber \\
&\times&\left[2q\mathrm{ln}q+(1+2q)(1-q)
+\frac{(\Gamma_eq)^2(1-q)}{2(1+\Gamma_eq)}\right],
\end{eqnarray}
where $\Gamma_e=4E_{IR}\Gamma/m_ec^2$, $q=E_1/\Gamma_e(1-E_1)$ and
$E_1=E_{\gamma}/\Gamma_e m_ec^2$. From IR/Optical (Mignani \& Caraveo 2001; 
Shibanov et al. 2003) and UV (Romani, Kargaltsev \& Pavlov 2005) 
observations, we adopt the following power low IR spectrum
\begin{equation}
\frac{dN_{IR}}{dE_{IR}}=N_{IR}\left(\frac{E_{IR}}{0.1\mathrm{eV}}\right)^{-1} 
\mathrm{~/cm^{3} erg}, 
\end{equation}
with  $N_{IR}=1.3\times10^{24}(d/0.25\mathrm{kpc})^2$. Integrating the
upscattering over the high-energy particles both inside and
outside the gap, we obtain the TeV inverse  Compton spectrum. 

\begin{figure}
 \begin{center}
\includegraphics[width=12cm, height=8cm]{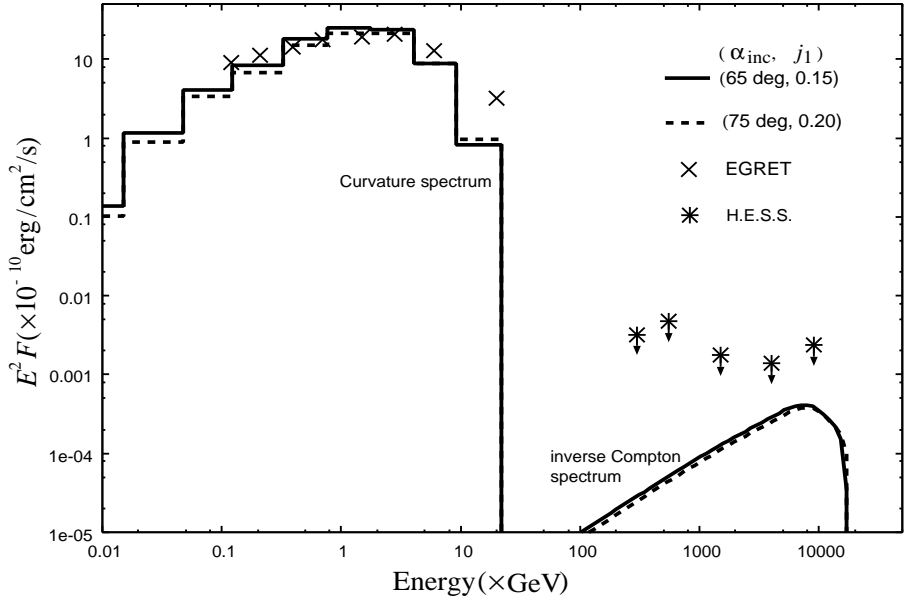}
\caption{Pulsed gamma-ray spectrum from the Vela pulsar.  The solid
line and dotted line show the spectrum for
$(\alpha_{inc},~j_1)=(65^{\circ},0.15)$ and $(75^{\circ},0.2)$,
respectively, with $(j_2,j_{gap})=(0,0.1)$. The upper limits in
0.1-1TeV range has been determined by H.E.S.S. (Konopelko et al. 2005).}
\label{invspec}
\end{center}
\end{figure}

Fig.\ref{invspec} shows the expected inverse Compton spectrum with the
curvature spectrum. The solid line and dotted line show the spectrum for 
$(\alpha_{inc},~j_1)=(65^{\circ},0.15)$ and $(75^{\circ},0.2)$,
respectively, with $(j_2,j_{gap})=(0,0.1)$. We find that 
the model TeV flux  less than $0.01\%$ of the GeV flux  is consistent
with the recent upper limits by H.E.S.S.(Konopelko et al. 2005). 
In the present model, the high-energy particles accelerated in the gap
make both the curvature and inverse-Compton spectra.
For such a case,  the flux ratio between the inverse-Compton  
and curvature processes  does not depend on the inclination angle 
and the gap size, because  
the ratio is determined by ratio between emission rates of a particle so
that 
\begin{eqnarray}
\frac{L_{IC}}{L_{curv}}&\sim&\frac{\Gamma m_ec^2\sigma_T
E_{IR}\frac{dN_{IR}}
{dE_{IR}}}{\frac{2}{3}\frac{e^2\Gamma^4}{R_c^2}} \nonumber \\
&=&6.4\times10^{-5}\left(\frac{\Gamma}{10^7}\right)^{-3}
\left(\frac{R_c}{\varpi_{lc}}\right)^2\left(\frac{E_{IR}dN_{IR}/dE_{IR}}
{10^{11}\mathrm{cm^{-3}}}\right).
\end{eqnarray}
It is clear  that the inverse-Compton process is less important for the
gap electrodynamics.

\subsection{Comparison with the previous works}
\label{compari}
We have estimated the inclination angle that $\alpha_{inc}\ge 65$$^{\circ}$ for
the Vela pulsar. This result is 
marginally  consistent with the result by RY95,
in which $\alpha_{inc}\sim 65$$^{\circ}$ for the Vela pulsar was predicted to 
explain the phase difference between the radio and $\gamma$-ray pulses.  
RY95 considered the gap having no
external component ($j_1=j_2=0$) and starting from the conventional null
surface. However, the present model has predicted the particle 
injection at the inner boundary ($j_1\neq0,j_2=0$) 
to explain the observed flux, although 
the obtained inner boundary is located near the conventional null 
surface (Fig.\ref{allgap}). 

RY95 have assumed that the gap is extending to the light cylinder
to produce the first peak of the light curves. In this paper, 
the gap geometry in Fig.\ref{allgap} that is favored for matching 
the spectrum of the Vela
pulsar  does not predict such long outer gap. 
However, we note that the accelerated particles 
in the outer gap emit the $\gamma$-ray
photons outside the gap. As  Fig.\ref{spectra} shows,  
the curvature photons emitted outside the gap 
by the outwardly moving particles also contribute to emissions 
above 100MeV. Specifically, the outwardly moving particles 
escape from the light
cylinder with the Lorentz factor $\Gamma\sim 1.2\times
10^{7}(\Omega/100\mathrm{s^{-1}})^{-1/3}(<R_c>/\varpi_{lc})^{2/3}$, the
corresponding  curvature photon energy is 
$E^{lc}_{c}\sim 160(<R_c>/\varpi_{lc})$~MeV, where $<R_c>$ is the
averaged curvature radius of the field line, along which the particles
move. Therefore, although  the outer boundary of the gap is located
inside of the light cylinder, the 
high-energy photons are emitted near the light cylinder and 
 the double peak structure in the calculated light curves is
expected. The detail calculation will be
done in the subsequent papers. 

Dyks \& Rudak (2003) proposed the radiation emission region starting
from near the stellar surface to explain the presence of 
the outer-wing emissions and of the off-pulse emissions of the Crab
pulsar. With  the two-dimensional
outer gap model, we found that such gap has a large 
particle  injection  at the outer boundary (Paper2).
 For such cases, $j_1\ll j_2$, however,
 the our model (e.g. \case~4) also predicts the 
softer spectrum than the observed one of the Vela pulsar,
as we discussed in \S\S\ref{diagnosis}.

\subsection{Screening of the electric field}
\label{inpart} 

As we have seen, the inner boundary, where $\tilde{E}_{||}=0$,
 is located at the position where the local charge density caused 
by the current is equal to the  Goldreich-Julian charge density, 
that is, $(j_1-j_2+j_g)\tilde{B}\sim\tilde{B_z}$ is satisfied on the
inner boundary. 
 However, the condition $|-(j_1-j_2+j_g)\tilde{B}|<|-\tilde{B}_z|$ 
is satisfied between the stellar surface and the inner boundary. 
This negative charge depletion from the Goldreich-Julian charge 
density predicts  a field-aligned electric field between the stellar
surface, on which $\tilde{\Phi}_{nco}=0$ is defined, and the inner
boundary. In such a case, there is a potential drop along the magnetic
field lines.  In the present paper,
however,  we have imposed the condition (\ref{cond1}),
$\tilde{\Phi}_{nco}=0$,  at the inner boundary 
with the assumption that a screening mechanism of the electric field 
 is working and that  the potential drop between the stellar surface
and the outer gap is much smaller than the potential drop in whole outer gap. 
In this subsection, therefore, we discuss the screening 
mechanism  between the stellar surface and the inner boundary with  
the case that $j_1=j_2=0$. 

Firstly, as the particles that screen the electric field, we mention a 
 population of the electrons trapped between 
the stellar surface and the outer gap,  such as the electron clouds
trapped between the stellar surface and  the boundary of the outer
magnetospheric gap  appeared in Smith, Michel \&  Thacker (2001).  
These electrons stabilize the force free
surface (i.e. the inner boundary)  between the plasma side and the
outer gap side. Therefore, we expect that three kind of the particles
will populate between the star and the inner boundary; 
the primary particles accelerated in the gap, the  pair particles produced 
outside gap by the inwardly $\gamma$-rays and the trapped electrons.
  The vacuum outer gap model of CHR86 also assumed
these trapped electrons between the stellar surface and the
traditional null charge surface.  
Such trapped electrons will be distributed to fill the negative
charge depletion, $j_g\tilde{B}-B_z$, from the Goldreich-Julian charge
density,  and to satisfy the condition $\tilde{E}_{||}=0$ between the
stellar surface and the inner boundary.

Because the inner boundary will be affected by  the
plasma condition around the gap, we consider a particular case 
that there are no trapped electrons as an another possibility, 
 that is, only primary and pair particles are
migrating toward the stellar surface along the magnetic field line linking 
the star and the outer gap.  With  the assumption  that 
the screening distance along the field line is much smaller 
than the trans-field thickness, we take the field-aligned part 
in the Poisson equation to obtain the field-aligned electric field. 
In this case, we can produce the solution that 
the field-aligned electric field $\tilde{E}_{||}$ undergoes spatial 
oscillation, and therefore is screened out by the small
velocity difference of  the pair particles. Because the amplitude of the
potential ($\sim 10^3$ in the non-dimensional value) is much smaller than
the  potential drop ($\sim 10^7$)  in the outer gap, the condition
(\ref{cond1}),  $\tilde{\Phi}_{nco}=0$,  at the inner boundary 
is a good treatment. Following Shibata, Miyazaki \& Takahara (2002), 
however, we can show  that a surprisingly large pair-creation
rate is required to screen. In fact, 
the screening is achieved if the pairs were created with a rate 
\[
\left<\frac{dm}{ds_{||}}\right>\geq
0.0025\left(\frac{\Omega}{100\mathrm{rads^{-1}}}\right)\left(\frac{\gamma_0}{10^3}\right)^2
\left(\frac{0.1\varpi_{lc}}{r}\right) ~\mathrm{/cm}, 
\]
where $m$ is the pair flux divided by $\Omega B_{in}/2\pi$, 
$B_{in}$ is the strength of the magnetic field at the inner boundary 
and $\gamma_0$ is the initial Lorentz factor of the pair.
This rate indicates that the total number $N_p$ of the pairs
produced within the typical screening distance $\sim 10^5$cm is 
$N_p/N_{GJ}\sim 10^2$, where $N_{GJ}$ represents 
the typical Goldreich-Julian value. Such a very large pair creation rate
has not been predicted by the present cascade model, in which
$dm/ds_{||}\sim 10^{-8}$~/cm (Fig.\ref{map}). On these ground, we
expect that the pair polarization does not screen out 
the field-aligned electric field within the small distance along 
the field line.

As discussed above, if no trapped electrons exist between the stellar
surface and the outer gap, the field-aligned electric field will be 
 caused outside the gap over a wide range  along the field line.
 For such a case,  the trans-field part in the Poisson equation determines the
non-corotational  potential, and the field-aligned part is not
important, so that  
\begin{equation}
\frac{\partial^2\tilde{\Phi}_{nco}}{\partial
\xi_{\perp}^2}\sim-[-\tilde{\rho}_0+
\tilde{\rho}_++\tilde{\rho}_-],
\label{transeq}
\end{equation}
where $\tilde{\rho}_0=j_g\tilde{B}-\tilde{B}_z$, $\tilde{\rho}_+$ and 
$\tilde{\rho}_-$ are, respectively, the charge densities of the par
positrons and electrons, and $\xi_{\perp}$ is the distance in the
trans-field direction. By ignoring the trans-field distribution 
of the charge density, the solution of equation (\ref{transeq}) becomes
$\tilde{\Phi}_{nco}\sim[-\tilde{\rho}_{0}+\tilde{\rho}_++\tilde{\rho}_-]
\xi_{\perp}
(\tilde{D}_{\perp}-\xi_{\perp})/2$,
where we use the boundary conditions that $\tilde{\Phi}_{nco}=0$ at
$\xi_{\perp}=0$ and $\xi_{\perp}=\tilde{D}_{\perp}$, where 
 $\tilde{D}_{\perp}$ represents the trans-field thickness. The field-aligned
electric field is
\begin{equation}
\tilde{E}_{||}=\frac{\partial\tilde{\Phi}_{nco}}{\partial\xi_{||}}\sim
\frac{\partial}{\partial
\xi_{||}}(-\tilde{\rho_{0}}+\tilde{\rho}_++\tilde{\rho}_-)
\frac{\xi_{\perp}(\tilde{D}_{\perp}-\xi_{\perp})}{2},
\label{twoelectric}
\end{equation}
where $\xi_{||}$ is the field-aligned distance increasing toward the star.
 
Let us consider the sign of the field-aligned electric field
 of equation (\ref{twoelectric}). If we ignore the contribution of the pair
particles on the electric field, we find that the
electric field becomes a positive because the outer gap is located on 
the field lines that curve away from the rotational axis
 and because the derivative $-\partial\tilde{\rho}_{0}/
\partial{\xi_{||}}$ has a positive value between the 
 stellar surface and the inner boundary.  
This positive electric field accelerates the pair electrons
 and decelerates the pair positrons, which are migrating toward the
star. The continuity equations of the
pair particles  predict  the field-aligned 
distribution of the charge density increasing toward the stellar surface, 
that is $\partial(\tilde{\rho}_{+}+\tilde{\rho}_-)/\partial{\xi_{||}}>0$. 
The derivative of the total charge density in the right
hand side of equation (\ref{twoelectric}) has positive value, 
and therefore the effect of the pair polarization strengthens 
the positive electric field, which will appear unless there 
are such pair particles. 
Although some positrons are returned and are accelerated toward the
gap by the positive electric field, the positive electric field will be  held
between the stellar surface and the inner boundary even
if the effect of the discharge of the pairs is taken into account, as
has been indicated by Hirotani (2005, in preparation). 

 Recently,  Hirotani (2005) has solved the 
 structure of the particular outer gap with the particle motion by
setting the inner boundary at the stellar surface.  He has
 indicated that such outer gap is composed of two region; one region
(called "minor part") takes on  a minor part of  the 
 potential drop of  whole outer gap and has a small positive field-aligned 
electric field, and  other (called "major part") takes on 
 a major part of the potential drop  and  has a large  field-aligned
electric field. The major part, which corresponds to so-called
outer gap in the present paper,  is located in the outer magnetosphere and
the minor part is extending between the stellar surface and the major
part.  Because the potential drop in the minor part of the gap 
is negligibly small in comparison with that in the major part, 
the boundary condition (\ref{cond1}),  $\Phi_{nco}=0$ (\ref{cond1}),  
at the inner boundary of the major gap will be a good treatment as long
as we are concerned with the major part of the gap. 

With above discussions, therefore, we conclude that 
the boundary condition that $\tilde{\Phi}_{nco}=0$ at the inner
boundary, where $\tilde{E}_{||}=0$, for the outer gap  is a good treatment.
\subsection{The connection between the global and local models} 
\begin{figure}      
 \begin{center}       
\includegraphics[width=12cm, height=8cm]{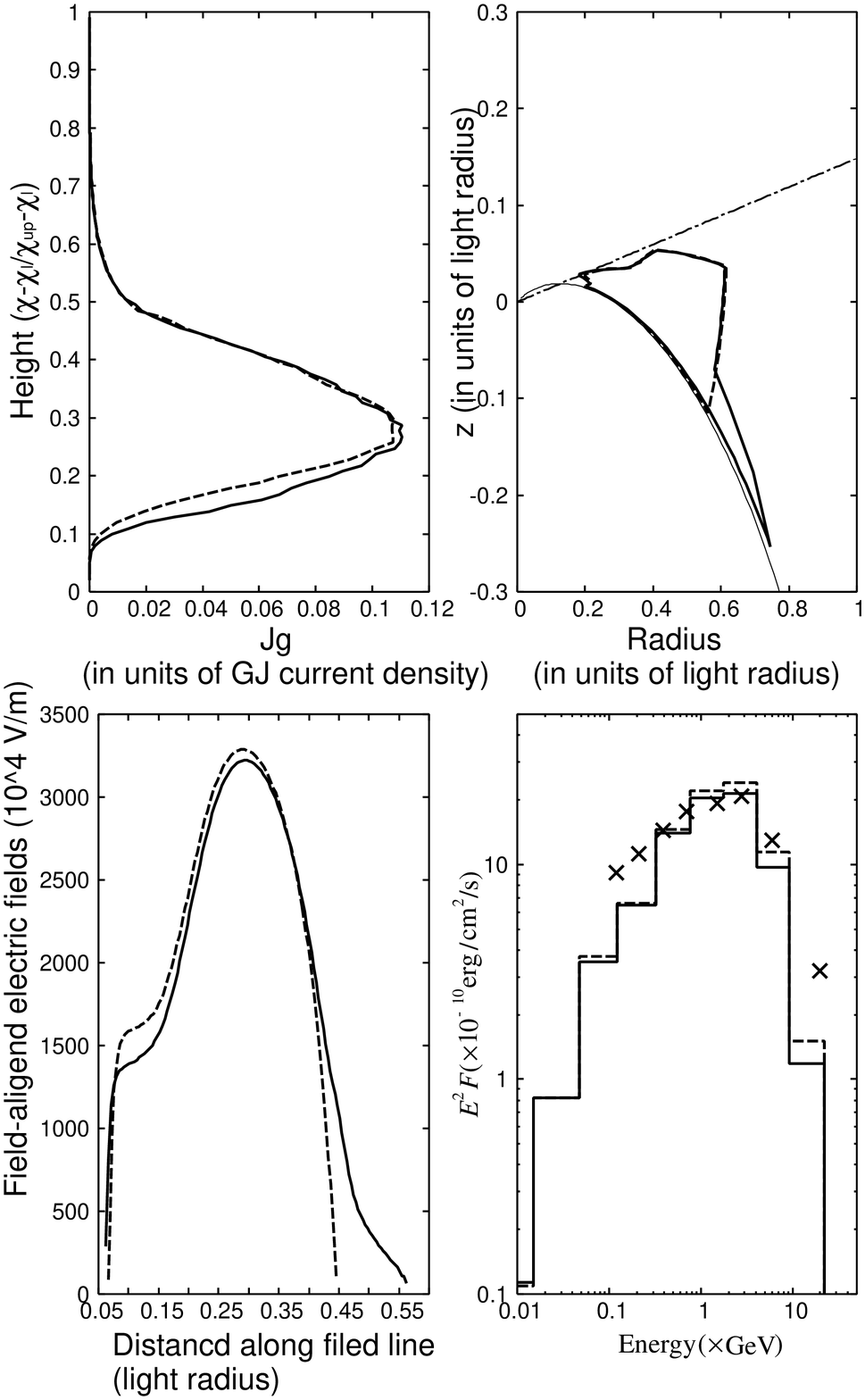}       
\caption{Outer gap structure with $(j_1,j_2)$=(0.1,0) and  
$\alpha_{inc}=65$$^{\circ}$. Each panel is same with Fig.\ref{gapcom}. 
The solid lines show the results of the gap having the outer boundary
located near the light cylinder at the lower part. The dashed lines
show the results of the gap having the outer boundary 
of $\zeta_{out}=0.64.$ }       
\label{speccom}      
\end{center}       
\end{figure}   

As we have seen,  the outer gap structure and produced $\gamma$-ray
spectrum are  controlled by the current, which circulates in 
the magnetosphere globally. 
By comparing the observation, our local model can 
 constrain the current flow pattern in the gap  and the inclination 
angle, which cannot be observed directly. For further discussions, 
we would need a connection between  the global and local models, as follows. 

In the previous  sections, we had considered the outer boundary following   
 a curve line perpendicular to the magnetic field lines, that
is $\zeta_{out}=$constant. It is notable that the shape of the outer
boundary  is controlled by the 
 trans-field distribution of  the total current.  

Around the lower boundary,  only a few pairs are produced by the pair creation,
 because the $\gamma$-rays are always emitted to the 
convex side of the magnetic field lines. Therefore, the shape of the
outer boundary can become such as the gap structure displayed with solid
lines in Fig.\ref{speccom}, of which the outer boundary changes from
$\zeta_{out}=0.64$ to $\zeta_{out}=1$ in region within 16\%  of 
 gap thickness measured from the lower boundary. For comparison, we show 
the gap structure (dotted line) having the outer boundary of 
$\zeta_{out}=0.64$. 

The  electrons produced in the extending region emit the inwardly
propagating $\gamma$-rays to convex side of the magnetic field lines
and make pairs around the inner boundary.  
In the upper left panel of Fig.\ref{speccom}, we can see that 
an effect of these new born pairs is the increase  of the internal 
current around the hight 
$(\chi-\chi_l)/(\chi_{up}-\chi_l)\sim 0.2$ measured from the lower boundary.
Fig.\ref{speccom}. The difference between the two currents
 produces the difference between  the outer boundaries 
in the upper right panel.
 
We find in Fig.\ref{speccom} that the calculated 
spectra for both cases are consistent with the observation.  
Therefore, it is difficult to deduce the shape of the outer boundary
 using the present local model. 
As a global study, CHR86 postulated that the outer gap is extending 
until the light cylinder, such as the gap geometry depicted with  
the solid line in Fig.\ref{speccom},  due to the  presence of 
an ionic (or a positronic)  particle outflow from the light cylinder.   
For the aligned-rotator, on the other hand,  Mestel et al. (1987) studied 
a structure of the magnetosphere with a global current and an acceleration region, which
closes within  the light cylinder (see also Mestel 1999).  
These global studies solve 
the global current, which interacts with the polar cap, 
the outer gap and the pulsar wind.  

Because the local model can predict the current through the gap 
with the observations as we have seen in the present paper,  
and because the global study can solve the global current 
flow  in the magnetosphere, the combination 
between the local and global studies 
would give a consistent  understanding of the structure 
of the active pulsar  magnetosphere. For such a discussion, 
a three-dimensional electrodynamical model is needed to 
compare with the observed phase-resolved spectra and light curves.

\section*{Acknowledgments}
The authors thank D.Melrose, K.S.Cheng and M.Takizawa for much
valuable discussion, and the anonymous referee for his/her  helpful
comments on improvements to the paper. This work was supported 
by the Theoretical Institute for 
Advanced Research in Astrophysics (TIARA) operated under Academia
Sinica and the National Science Council Excellence Projects program 
in Taiwan administered through grant number NSC 94-2752-M-007-001 
and NSC 94-2752-M-007-002. 

%%%%%%%%%%%%%%%%%%%%%%%%%%%%%%%%%%%%%%%%%%%%%%%%%%%%%%%%%%%%%%%%

%%%%%%%%%%%%%%%%%%%%%%%%%%%%%%%%%%%%%%%%%%%%%%%%%%%%%%%%%%%%%
%\label{lastpage}

\begin{thebibliography}{}
\bibitem[\protect\citeauthoryear{Arons}{1983}]{arons}
Arons J., 1983, ApJ, 266, 215
\bibitem[\protect\citeauthoryear{Blumenthal}{1970}]{blume}
Blumenthal G.R., Gould R.J., 1970, Rev. Mod. Phys., 42, 237
\bibitem[\protect\citeauthoryear{Cha, Semback \& Danks}{1999}]{CSD}
Cha A.N., Sembach K.R., Danks A.C. 1999, ApJ, 515L, 25
\bibitem[\protect\citeauthoryear{Cheng, Ho \& Ruderman}{1986}]{CHRa}
  Cheng K.S., Ho C., Ruderman M., 1986a, ApJ, 300, 500 (CHR86)
\bibitem[\protect\citeauthoryear{Cheng, Ho \& Ruderman}{1986}]{CHRb}
  Cheng K.S., Ho C., Ruderman M., 1986b, ApJ, 300, 522
\bibitem[\protect\citeauthoryear{Cheng, Ruderman \& Zhang}{2000}]{CRZ}
  Cheng K.S., Ruderman M., Zhang L., 2000, ApJ, 537, 964
\bibitem[\protect\citeauthoryear{Daugherty \& Harding}{1996}]{DH}
  Daugherty J.K., Harding A.K., 1996, ApJ, 458, 278
\bibitem[\protect\citeauthoryear{Dyks \& Rudak}{2003}]{DR}
  Dyks J., Rudak B., 2003, ApJ. 598, 1201
\bibitem[\protect\citeauthoryear{Dyks, Harding \& Rudak}{2004}]{DHR}
  Dyks J., Harding A.K.,  Rudak B., 2004, ApJ, 606, 1125
\bibitem[\protect\citeauthoryear{Fierro, Michelson, Nolan \& Thompson}{1998}]{FMNT}
Fierro J.M., Michelson P.F., Nolan P.L., Thompson D.J. 1998, ApJ, 494,
734
\bibitem[\protect\citeauthoryear{Goldreich \& Julian}{1969}]{GJ}
Goldreich P., Julian W.H., 1969, ApJ, 157, 869
\bibitem[\protect\citeauthoryear{Harding, Strickman, Gwinn, Dodson,
Moffet, \& McCulloch}{2002}]{HSGDMM}
Harding A.K., Strickman M.S., Gwinn C., Dodson R., Moffet D., 
McCulloch P., 2002, ApJ,  576, 376
\bibitem[\protect\citeauthoryear{Hirotani}{2005}]{H}
  Hirotani K., 2005, in  preparation
\bibitem[\protect\citeauthoryear{Hirotani \& Shibata}{1999a}]{HS1a}
  Hirotani K., Shibata S., 1999a, MNRAS, 308, 54 (HS99)
\bibitem[\protect\citeauthoryear{Hirotani \& Shibata}{1999b}]{HS1b}
  Hirotani K., Shibata S., 1999b, MNRAS, 308, 67
\bibitem[\protect\citeauthoryear{Hirotani \& Shibata}{2001a}]{HS2a}
  Hirotani K., Shibata S., 2001a, MNRAS, 325, 1228
\bibitem[\protect\citeauthoryear{Hirotani \& Shibata}{2001b}]{HS2b}
  Hirotani K., Shibata S., 2001b, ApJ, 558, 216	
\bibitem[\protect\citeauthoryear{Hirotani, Hardin \& Shibata}{2003}]{HHS}
  Hirotani K., Harding A.K., Shibata S., 2003, ApJ, 591, 334
\bibitem[\protect\citeauthoryear{Kanbach, et al.}{1994}]{Kan}
Kanbach G. et al. 1994, A\&A, 289, 855 
\bibitem[\protect\citeauthoryear{Konopelko, et al.}{2005}]{Kono}
Konopelko A. et al., 2005, 29th International Cosmic Ray Conference
Pune, 101-106
\bibitem[\protect\citeauthoryear{Mestel}{1999}]{Mestel}
Mestel L., 2003, Stellar Magnetism, International Series of Monographs of 
Physics. Oxford Univ. Press, Oxford
\bibitem[\protect\citeauthoryear{Mestel Robertson Wang \&
Westfold}{1985}]{MRWW}
Mestel L., Robertson J.A., Wang Y.-M., Westfold K.C., 1985, MNRAS, 217, 443
\bibitem[\protect\citeauthoryear{Mignani}{2001}]{Mignani}
Mignani R.P., Caraveo P.A. 2001, A\&A, 376, 213
\bibitem[\protect\citeauthoryear{Muslimov\& Hardin}{2004}]{MH}
  Muslimov, A.G.,  Harding A.K., 2004, ApJ, 606, 1143
\bibitem[\protect\citeauthoryear{\"{O}gelman et al.}{1993}]{OFZ}
\"{O}gelman H., Finley J.P., Zimmerman H.U., 1993, Nat, 361, 136
\bibitem[\protect\citeauthoryear{Press et al.}{1988}]{PFTV}
Press W.H., Flannery B.P., Teukolsky S.A., Vetterling W.T.,
 1988, Numerical Recipes (Cambridge: Cambridge Univ. Press)
\bibitem[\protect\citeauthoryear{Pavlov, Zavlin, Sanwal, Burwitz \& Garmire}{2001}]{PXDBG}
Pavlov G.G., Zavlin V.E., Sanwal D., Burwitz V., Garmire G.P., 2001, ApJ,
552, L129
\bibitem[\protect\citeauthoryear{Qial et al.}{2004}]{Q}
Qiao G.J., Lee K.J., Wang H.G., Xu R.X., Han J.L., 2004, ApJ, 606, L49
\bibitem[\protect\citeauthoryear{Romani}{1995}]{R}
  Romani R.W., 1996, ApJ, 470, 469
\bibitem[\protect\citeauthoryear{Romani et al}{2005}]{RKP}
  Romani R.W., Kargaltsev O., Pavlov G.G., 2005, ApJ, 627, 383
\bibitem[\protect\citeauthoryear{Romani \& Yadigaroglu}{1995}]{RY}
  Romani R.W., Yadigaroglu I.A., 1995, ApJ, 438, 314 (RY95)
\bibitem[\protect\citeauthoryear{Ruderman, \& Sutherland}{1975}]{RS}
  Ruderman M.A., Sutherland P.G., 1975, ApJ, 196, 51
\bibitem[\protect\citeauthoryear{Scharlemann, Arons \& Fawley}{1978}]{ASF}
Scharlemann E.T., Arons J., Fawley W.M., 1978, ApJ, 222. 297
\bibitem[\protect\citeauthoryear{Smith, Michel \&
Thacker}{2001}]{SMT}
Smith I.A., Michel F.C., Thacker P.D., 2001, MNRAS, 322. 209
\bibitem[\protect\citeauthoryear{Shibata}{1995}]{Shibata}
  Shibata S., 1995, MNRAS, 276, 537
\bibitem[\protect\citeauthoryear{Shibata}{2002}]{SMT}
  Shibata S., Miyazaki J., Takahara F., 2002, MNRAS, 336, 233
\bibitem[\protect\citeauthoryear{Shibanov}{2003}]{Shibanov}
  Shibanov Yu.A., Koptsevich A.B., Sollerman J., Lundquist P.,
2003, A\&A, 406, 645
\bibitem[\protect\citeauthoryear{Sturrock}{1971}]{Stu}
  Sturrock P.A., 1971, ApJ, 164, 529
\bibitem[\protect\citeauthoryear{Takata, Shibata \& Hirotani}{2004}]{TSH1}
  Takata J., Shibata S., Hirotani K., 2004a, MNRAS, 348, 241 (Paper1)
\bibitem[\protect\citeauthoryear{Takata, Shibata \& Hirotani}{2004}]{TSH2}
  Takata J., Shibata S., Hirotani K., 2004b, MNRAS, 354, 1120 (Paper2)
\bibitem[\protect\citeauthoryear{Thompson et al.}{1999}]{Th2}
  Thompson D.J. et al., 1999, ApJ, 516, 297
\bibitem[\protect\citeauthoryear{Zhang \& Cheng}{1997}]{ZC}
Zhang L., Cheng K.S., 1997, ApJ, 487. 370
\end{thebibliography}
\end{document}